\documentclass[preprint,12pt]{elsarticle}

\usepackage{amssymb}
\usepackage{amsmath}

\usepackage{tikz}
\usepackage{tikz}
\usetikzlibrary{
    arrows.meta,
    positioning,
    calc
}

\usepackage[caption=false]{subfig}

\usepackage{longtable}

\journal{eTransportation}

\begin{document}

\begin{frontmatter}

\title{Quantifying Realizable Flexibility Limits in Fast and Ultra-Fast EV Charging Using Real-World Data} 

\author[inst1]{Cesar Diaz-Londono \corref{cor}}
\author[inst1,inst2]{Liu Zhang}
\author[inst1]{Jorge De La Cruz}
\author[inst1,inst3]{Hamidreza Arasteh}
\author[inst1,inst4]{Anand R.}
\author[inst5]{Daogui Tang}
\author[inst1]{Josep M. Guerrero}

\cortext[cor]{Cesar Diaz-Londono, email: cesar.diaz@ieee.org}

\affiliation[inst1]{organization={Center for Research on Microgrids (CROM), Huanjiang Laboratory},
            city={Zhuji, Shaoxing, Zhejiang},
            postcode={311800}, 
            country={China}}
            
\affiliation[inst2]{organization={School of Aeronautics and Astronautics, Zhejiang University},
            city={Hangzhou, Zhejiang},
            postcode={310000}, 
            country={China}}

\affiliation[inst3]{organization={Power Systems Operation and Planning Research Department, Niroo Research Institute (NRI)},
            city={Tehran},
            country={Iran}}

\affiliation[inst4]{organization={Department of Electrical and Electronics Engineering, Amrita Vishwa Vidyapeetham},
            city={Bengaluru},
            country={India}}

\affiliation[inst5]{organization={School of Transportation and Logistics Engineering, Wuhan University of Technology},
            city={Wuhan},
            postcode={430063}, 
            country={China}}

\begin{abstract}

The rapid growth of electric vehicles (EVs) is increasing the need to accurately quantify their flexibility as a resource for power system operation. However, most existing approaches rely on simplified or power-controllable models that overlook the intrinsic constraints of fast and ultra-fast DC charging. In practice, flexibility is fundamentally shaped by battery management system (BMS) behavior, connection time availability, and battery-protection limits. This paper introduces a trajectory-aware data-driven framework to quantify EV charging flexibility as an energy-bounded and time-constrained process. Based on 252 real charging sessions, 141 representative Power--SoC profiles are reconstructed to capture real-world charging dynamics. Unidirectional flexibility is defined through bounds on the maximum shiftable charging energy, while bidirectional flexibility is quantified as the bounds of the maximum extractable discharge energy under feasibility constraints. Results show that flexibility depends on charging state and connection time. Charging beyond 80\% SoC increases duration with limited gains, while higher charger power saturates due to BMS limits. Charging time in the 20\%--80\% range drops by over 60\%, and mean power increases by up to 40\%. The maximum extractable bidirectional energy can exceed twice its value depending on the point at which flexibility is activated. These results highlight that EV flexibility is not a controllable resource, but a bounded and time-dependent capability. As such, the proposed framework provides actionable limits that can be directly used by system operators and aggregators for scheduling, peak shaving, and short-duration flexibility services.

\end{abstract}

\begin{graphicalabstract}
 \includegraphics[width=\linewidth]{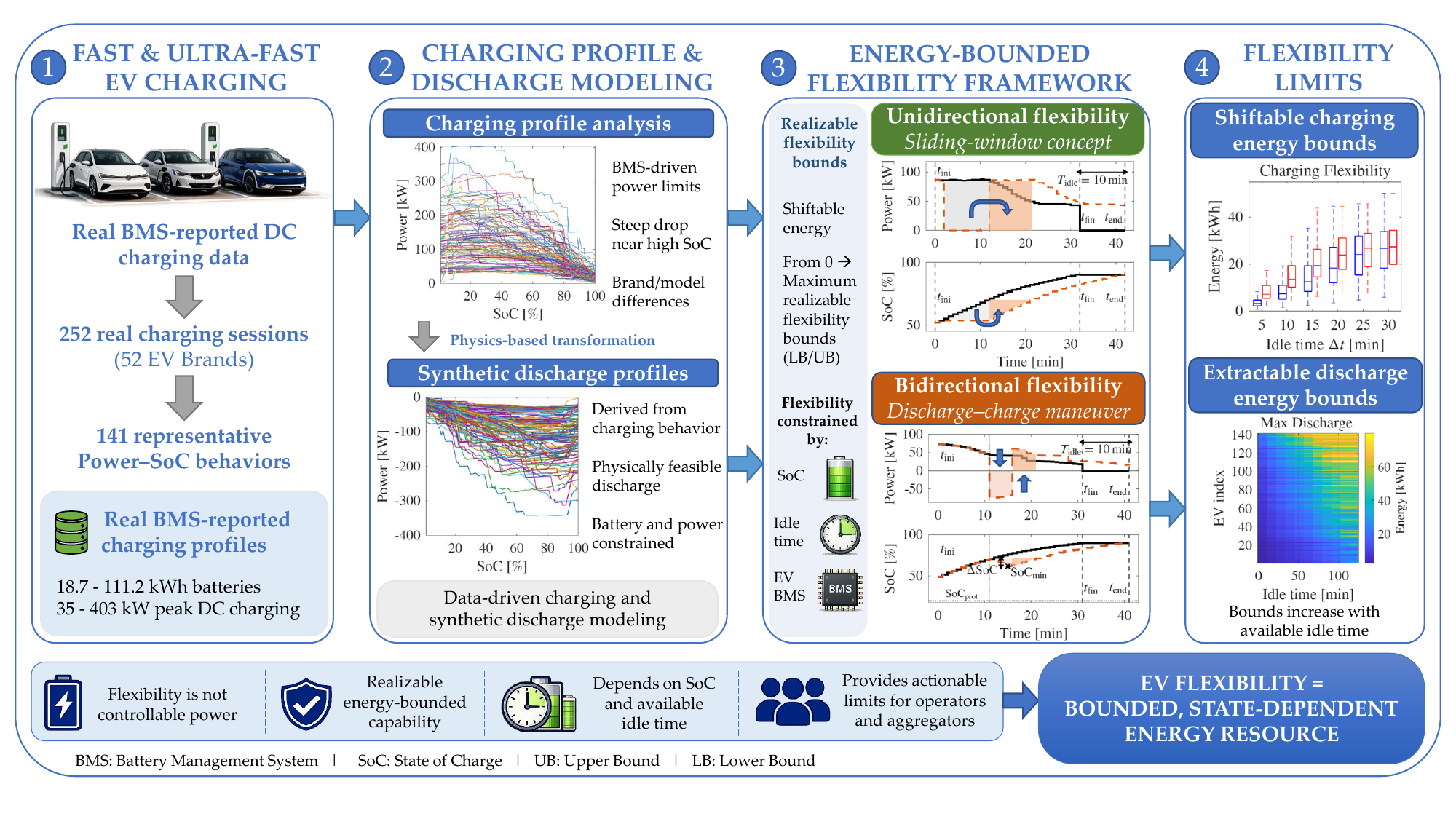}
\end{graphicalabstract}

\begin{highlights}
\item EV flexibility redefined as energy-bounded under BMS constraints
\item 141 real fast/ultra-fast EV charging profiles analyzed
\item Sliding-window reveals time-coupled limits of shiftable energy
\item Bidirectional flexibility bounded by time and battery constraints
\item Extractable energy can exceed twice depending on activation timing
\end{highlights}


\end{frontmatter}




\section{Introduction}

The global transition toward transportation electrification is accelerating rapidly as electric vehicles (EVs) become a key component of future low-carbon energy systems. Global EV sales exceeded 20 million vehicles in 2025, representing nearly one-quarter of all new car sales worldwide. In parallel, public charging infrastructure surpassed 5 million chargers globally, with ultra-fast charging (UFC) stations above 150~kW experiencing particularly strong growth~\cite{IEA2026}. According to the International Energy Agency (IEA), public charging infrastructure is commonly classified into slow charging ($<22$~kW), fast charging (22--150~kW), and ultra-fast charging ($>150$~kW), with the latter becoming increasingly important for long-distance and high-utilization applications~\cite{IEA2026}.

This rapid electrification process is occurring within the broader transformation of future power systems, commonly described through the ``5Ds'': decarbonization, decentralization, digitalization, democratization, and deregulation~\cite{Asghar2021}. In this context, EVs are increasingly viewed not only as transportation assets but also as flexible distributed energy resources capable of supporting renewable integration, congestion mitigation, ancillary services, and electricity market participation. Consequently, EV charging flexibility has emerged as a key enabler for the coordinated interaction between transportation and power systems.

At the same time, the integration of EVs into virtual power plant frameworks is transforming EV charging from an isolated optimization problem into a system-level coordination challenge involving forecasting, dispatch, and multi-agent decision-making under uncertainty~\cite{WANG2026}. Recent studies have explored the economic and operational value of EV flexibility for applications such as peak shaving, frequency regulation, energy trading, renewable-energy-following operation, and multi-service provision~\cite{Englberger2021,CHEMUDUPATY2025}. However, most existing approaches implicitly assume that EV flexibility is freely dispatchable, without explicitly characterizing the physical limits imposed by charging trajectories, idle-time availability, and battery-protection constraints.

Meanwhile, recent advances in UFC technologies have enabled charging powers above 350~kW, significantly reducing charging times and improving EV usability for long-distance and intensive-duty applications~\cite{Franzese2025}. Unlike conventional low-power charging, fast and ultra-fast DC charging operate under highly nonlinear and tightly constrained regimes, where charging power strongly depends on SoC, battery temperature, chemistry, and BMS actions. These operating conditions introduce additional challenges related to battery degradation, thermal stress, and grid impact~\cite{Franzese2025}. As a result, simplified constant-power charging assumptions commonly adopted in the literature are inadequate for the realistic characterization of EV charging flexibility.


\subsection{Related Work and Research Gap}

Multiple approaches have been proposed for EV flexibility characterization and aggregation. Data-driven and bottom-up methods have been used to estimate charging demand and derive flexibility bounds from mobility patterns and user behavior~\cite{Herrera2025}. Stochastic and probabilistic formulations have also been developed to quantify flexibility under uncertainty across different temporal and spatial scales~\cite{Tang2025,NUKIC2025}. More recently, deep-learning-based approaches have been introduced to forecast EV flexibility boundaries and operational uncertainty using large-scale charging datasets~\cite{WU2026}.

Long-term empirical analyses of charging trajectories have also been used to characterize EV charging behavior, charging regularity, and flexibility potential under realistic operating conditions~\cite{YU2025a}. Other studies incorporated real-world charging patterns to evaluate coordinated charging and load-shifting capabilities while highlighting the importance of idle time and user participation~\cite{LI2025}. In parallel, scalable aggregate flexibility formulations have been proposed for coordinated charging and grid-oriented optimization of large EV populations~\cite{LI2026}.

At the infrastructure and system level, flexibility regions and operational scheduling strategies have been investigated for charging stations considering bidirectional EV participation, user behavior, and energy management constraints~\cite{Zhang2025}. Aggregated EV flexibility has also been represented as equivalent energy storage resources to support flexibility quantification and grid integration analyses~\cite{CHEN2025e}. In particular, EV flexibility has been modeled as virtual storage or virtual grid capacity capable of supporting building-level flexibility management and deferring infrastructure investments~\cite{CHEN2025}. Other studies proposed convex and polyhedral representations of aggregate EV flexibility using operating envelopes and fleet-level flexibility sets for coordinated market participation and renewable integration~\cite{Lai2025,AlTaha2025}. These works demonstrate the increasing interest in exploiting EV flexibility for system-level coordination and infrastructure planning.

Despite these advances, many existing flexibility formulations are still based on simplified charging representations, particularly in the context of residential and slow-charging applications \cite{Diaz2023}. In many studies, charging power is assumed to be freely controllable within predefined limits, while flexibility is modeled through arbitrary temporal shifts, constant-power charging strategies, or generalized battery abstractions. Such assumptions become particularly problematic in fast and ultra-fast DC charging, where the charging process is tightly constrained by the BMS-governed Power--SoC relationship.

Recent studies have highlighted the importance of incorporating realistic charging curves and BMS-aware charging behavior into fast-charging coordination frameworks \cite{Saner2023,Saner2024IEEE}. These studies demonstrate the limitations of constant-power charging assumptions and highlight the importance of realistic charging trajectories in fast-charging systems. Similarly, recent flexibility aggregation studies have recognized the challenges associated with heterogeneous charging dynamics, scalability, and trajectory-based flexibility characterization \cite{LI2025e}. However, existing BMS-aware and aggregation-oriented approaches remain primarily focused on scheduling, optimization, or operational flexibility representations, without explicitly quantifying the physically achievable flexibility limits imposed by charging trajectories, idle-time availability, and battery-protection constraints.

As a result, EV flexibility should not be interpreted as a freely dispatchable resource, but rather as a bounded and condition-dependent capability. In practice, the physically achievable flexibility of an EV charging session is jointly determined by: (i) the EV-specific Power--SoC charging characteristics, (ii) the available connection time beyond the mandatory charging duration, and (iii) battery-protection limits restricting operating ranges and depth-of-discharge. Consequently, the practically relevant quantity is not arbitrary power modulation, but the \emph{maximum realizable flexibility} achievable under physical and operational constraints.


\subsection{Proposed Framework and Main Contributions}

To address the limitations identified in the previous subsection, this work proposes a data-driven framework for quantifying EV flexibility as an energy-bounded and time-constrained process directly derived from experimentally measured fast and ultra-fast charging trajectories. Unlike existing approaches based on generalized battery abstractions or arbitrary power-control assumptions, the proposed methodology preserves realistic BMS-aware Power--SoC dynamics and explicitly characterizes the physically achievable limits of both unidirectional and bidirectional flexibility.

Within the proposed framework, flexibility is quantified through lower and upper bounds defining a feasible flexibility region within which all admissible charging and discharging trajectories must lie. For unidirectional operation, flexibility is defined as the maximum shiftable charging energy achievable within finite time windows while preserving the mandatory charging trajectory. The adopted sliding-window formulation preserves the temporal continuity of the charging trajectory and directly quantifies the energy that can be shifted within finite user connection intervals. For bidirectional operation, flexibility is quantified as the maximum extractable discharge energy constrained by charging feasibility, battery-protection limits, and available idle time.

At the fleet level, the framework enables the construction of aggregate flexibility bounds, feasibility maps, and flexibility indicators that explicitly account for EV heterogeneity and idle-time availability. This provides a physically grounded basis for integrating EV flexibility into system-level studies, including congestion mitigation, renewable integration, peak shaving, and ancillary service provision.

To the best of the authors’ knowledge, this is the first study that analyzes hundreds of real fast- and ultra-fast DC charging profiles spanning multiple EV brands and platforms to derive a unified framework for unidirectional and bidirectional flexibility assessment under realistic BMS-aware charging behavior.

The main contributions of this work are summarized as follows:

\begin{itemize}

    \item \textit{Comprehensive BMS-aware dataset of fast and ultra-fast charging profiles.}  
    Based on 252 measured DC charging sessions, 141 distinct Power--SoC trajectories are extracted and mapped to brand--model groups. Complete 1--100\% charging profiles are reconstructed to enable consistent analysis across the EV fleet.

    \item \textit{Systematic evaluation of charging performance across charger ratings.}  
    Each EV is evaluated under multiple charger power levels (7--480~kW), quantifying charging time and average delivered power for both full charging cycles and practical operating intervals.

    \item \textit{Data-driven quantification of intra-session flexibility.}  
    Sliding-window metrics are introduced to estimate shiftable energy and normalized SoC flexibility over user-realistic time intervals, enabling a large-scale empirical assessment of flexibility potential in fast and ultra-fast charging systems.

    \item \textit{Bounded bidirectional flexibility assessment under limited V2G data availability.}  
    A data-driven framework is proposed to estimate upper and lower bounds of bidirectional flexibility using only experimentally measured charging sessions. By combining scaled symmetry-based discharging envelopes with window-based energy metrics, realistic and operationally relevant flexibility limits are obtained for planning and system-level studies.

\end{itemize}


\subsection{Structure of the Paper}

The remainder of this paper is organized as follows. Section~\ref{sec:EV_analysis} presents the EV--EVCS interaction framework, the adopted dataset, the reconstruction of complete BMS-aware charging profiles, and the generation of discharge trajectories for bidirectional analyses. Section~\ref{sec:flexibility} introduces the proposed flexibility formulation and defines the methodology used to quantify unidirectional and bidirectional flexibility at both single-EV and fleet levels. Section~\ref{sec:results} presents the charging performance analysis and the flexibility assessment results across different charger ratings and EV groups. Section~\ref{sec:discussion} discusses the implications of the proposed framework for grid operation, flexibility services, and system-level energy management. Finally, Section~\ref{sec:conclusions} summarizes the main findings and outlines directions for future work.


\section{Data and BMS-Aware Charging Behavior}
\label{sec:EV_analysis}

The charging performance of modern EVs at fast and ultra-fast DC charging stations is determined not only by charger capability, but often predominantly by the control logic implemented by the vehicle’s BMS. Consequently, understanding EV behavior under high-power charging conditions requires a detailed examination of the communication, control, and physical interactions that shape the delivered power trajectory. This section provides the empirical foundation of the study, including the EV--charger control mechanisms, the adopted dataset, the reconstruction of complete Power--SoC profiles, and the generation of consistent discharge profiles required for bidirectional flexibility analyses.


\subsection{EV--Charger Communication and Control Mechanisms}
\label{subsec:EVCS_comm}

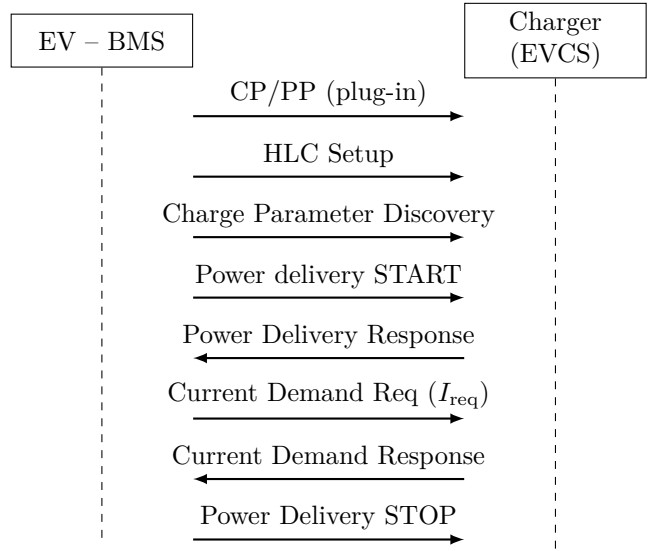
\begin{figure}[ht]
 \centering
 \footnotesize
 \begin{tikzpicture}[
     >=latex,
     font=\footnotesize,
     node distance=4cm,
     lifeline/.style={draw, rectangle, minimum width=2.4cm, minimum height=0.7cm, align=center},
     msg/.style={->, thick}
 ]

 \node[lifeline] (ev) {EV -- BMS};
 \node[lifeline, right of=ev, xshift=2.0cm] (evse) {Charger\\(EVCS)};

 \draw[dashed] (ev.south) -- ++(0,-6.3);
 \draw[dashed] (evse.south) -- ++(0,-6.3);

 \def\yA{-1.0}
 \def\yB{-1.8}
 \def\yC{-2.6}
 \def\yD{-3.4}
 \def\yE{-4.2}
 \def\yF{-5.0}
 \def\yG{-5.8}
 \def\yH{-6.6}

 \draw[msg] ([yshift=\yA cm]ev.east) -- 
            node[above,pos=0.5]{CP/PP (plug-in)} 
            ([yshift=\yA cm]evse.west);

 \draw[msg] ([yshift=\yB cm]ev.east) -- 
            node[above,pos=0.5]{HLC Setup} 
            ([yshift=\yB cm]evse.west);

 \draw[msg] ([yshift=\yC cm]ev.east) -- 
            node[above,pos=0.5]{Charge Parameter Discovery} 
            ([yshift=\yC cm]evse.west);

 \draw[msg] ([yshift=\yD cm]ev.east) -- 
            node[above,pos=0.5]{Power delivery START} 
            ([yshift=\yD cm]evse.west);

 \draw[msg] ([yshift=\yE cm]evse.west) -- 
            node[above,pos=0.5]{Power Delivery Response} 
            ([yshift=\yE cm]ev.east);

 \draw[msg] ([yshift=\yF cm]ev.east) -- 
            node[above,pos=0.5]{Current Demand Req ($I_{\mathrm{req}}$)} 
            ([yshift=\yF cm]evse.west);

 \draw[msg] ([yshift=\yG cm]evse.west) -- 
            node[above,pos=0.5]{Current Demand Response} 
            ([yshift=\yG cm]ev.east);

 \draw[msg] ([yshift=\yH cm]ev.east) -- 
            node[above,pos=0.5]{Power Delivery STOP} 
            ([yshift=\yH cm]evse.west);

 \end{tikzpicture}

 \caption{Simplified EV--EVCS communication sequence for charging (ISO~15118).}
 \label{fig:ev_EVCS_comm}
 \end{figure}

The charging behavior in EV charging sessions is determined by the interaction between the EV and the charging station (EVCS). This interaction can use the ISO~15118 communication framework~\cite{Shin2016}, following the simplified message exchange shown in \figurename~\ref{fig:ev_EVCS_comm}.

The session starts with the physical plug-in, where the Control Pilot (CP) and Proximity Pilot (PP) signals confirm connection and the admissible charging mode. Then, the High-Level Communication (HLC) channel is established to enable secure EV--EVCS data exchange. During Charge Parameter Discovery, the EV reports its battery limits and the EVCS returns its operational constraints, defining the feasible charging envelope. Once Power Delivery is enabled, the control loop is driven by Current Demand messages: at each cycle, the BMS requests a charging current $I_{\mathrm{req}}$ according to its internal strategy and safety limits, while the EVCS enforces hardware bounds and replies with the applied voltage and delivered current.

The resulting DC charging power is:
\begin{equation}
    P(k) \approx V_{\mathrm{bat}}(k)\, I_{\mathrm{req}}(k),
\end{equation}
with the EVCS operating as a controlled DC converter that adjusts its output voltage and current within hardware limits while following the charging requests and battery constraints communicated by the BMS. Consequently, the BMS governs the overall evolution of the charging-power trajectory, while the EVCS ensures safe operation and compliance with electrical constraints. When the EV reaches its target \(\mathrm{SoC}\) or charging termination condition, it issues a PowerDelivery(STOP) command. The EVCS then ramps down the current, ceases power transfer, and closes the session, completing the charging process.


\subsection{Dataset Description and EV Technical Specifications} \label{subsec:data}

Since the EV BMS governs the evolution of the charging-power trajectory, empirical datasets are essential to characterize how different EV models behave under real fast- and ultra-fast-charging conditions. To that end, this work relies on the real BMS-reported charging profiles published by Fastned~\cite{Fastned2025}, which represent one of the most comprehensive public datasets of DC charging curves available today. This public dataset contains 252 measured charging sessions, each corresponding to an individual EV, from which 141 distinct charging behaviours were extracted based on their Power--SoC trajectories (see \ref{app1} for the list of evaluated EVs). Each distinct behaviour corresponds either to a single EV model or to a group of EVs that exhibit identical charging patterns, reflecting the model-specific implementation of the BMS charging strategy described earlier.

\begin{figure}
     \centering
     \includegraphics[width=1.34\linewidth, angle=-90]{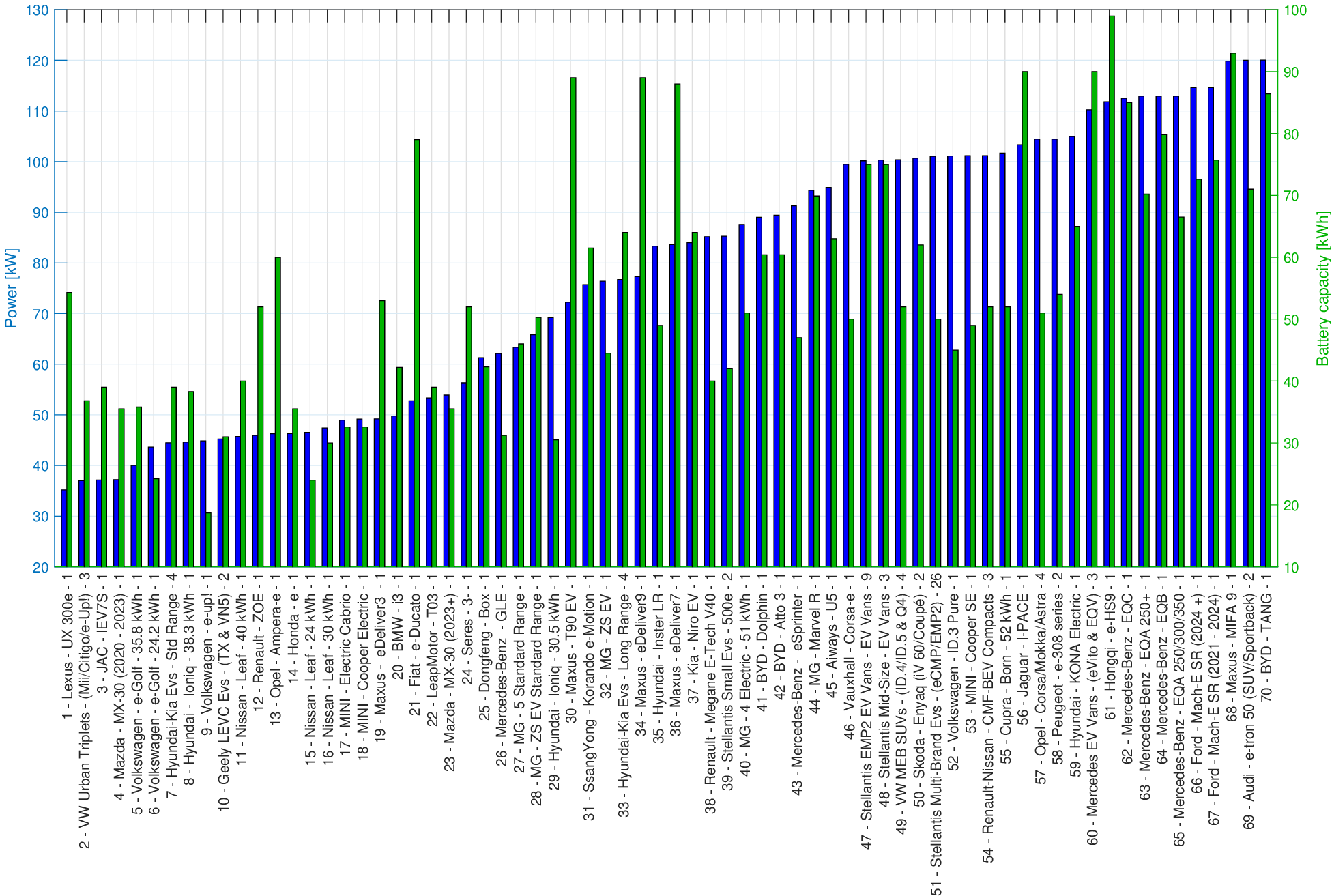}
     \caption{EVs with IDs 1--70: battery capacity (green) and maximum absorbed DC charging power (blue). Labels follow the format “ID -- Brand/Model -- Count”, where Count indicates how many EVs share the same charging behaviour.}
     \label{fig:EVbrands1}
 \end{figure}

\begin{figure}
     \centering
     \includegraphics[width=1.38\linewidth, angle=-90]{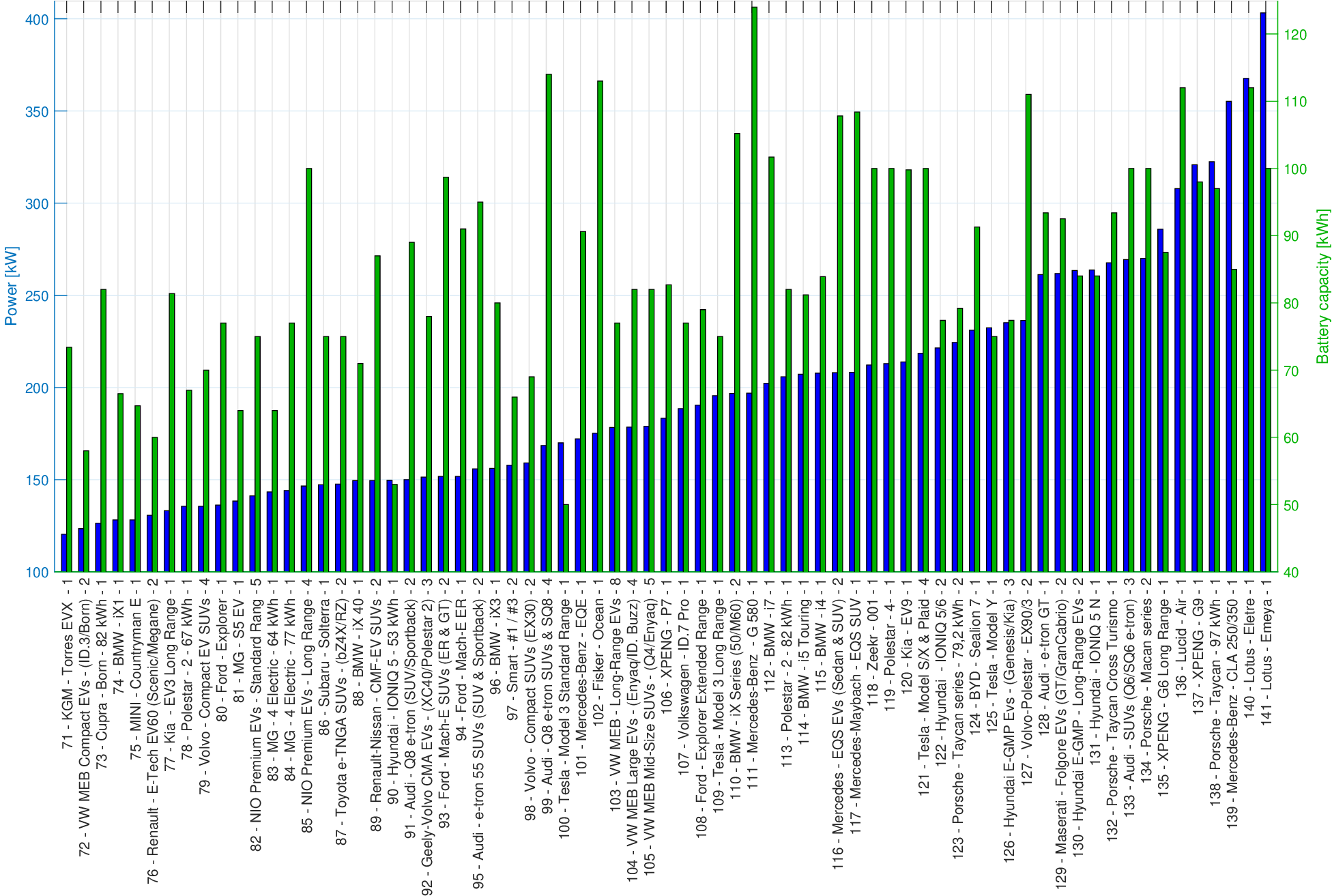}
     \caption{EVs with IDs 71--141: continuation of the catalogue of distinct charging behaviours, showing battery capacity (green) and peak charging power (blue).}
     \label{fig:EVbrands2}
 \end{figure}

\figurename s~\ref{fig:EVbrands1} and~\ref{fig:EVbrands2} summarize these 141 unique EV profiles. EVs are sorted by their maximum absorbed DC charging power, and each entry is represented by an ``EV ID'' used throughout the analysis.  

For each EV ID, two quantities are shown:
\begin{itemize}
    \item Peak charging power requested by the BMS during the charging session (blue).
    \item Nominal battery capacity (green).
\end{itemize}

For each bar, the label follows the format $\text{ID -- Brand/Model -- Count}$, where \textit{Count} indicates how many EVs share the same charging behaviour. For example, the entry “51 -- Stellantis Multi-Brand EVs - (eCMP/EMP2) -- 26’’ indicates that 26 EV models (listed in \ref{app1}) behave identically in their Power–SoC profile, despite belonging to different brands and body types, reflecting the adoption of a common battery platform and BMS charging logic.

The figures highlight the wide heterogeneity of the modern EV fleet. The smallest reported battery capacity is 18.7~kWh, whereas the largest reaches 111.2~kWh. In terms of peak charging power, the minimum observed value is 35~kW and the maximum reaches 403~kW for flagship ultra-fast charging models. These variations show that flexibility can differ substantially across EVs.


\subsection{EV Charging Profiles from Measured Data} \label{subsec:EVprof}

The dataset reveals a wide diversity of EV models, battery capacities, and peak
DC charging powers. However, the raw Power--SoC curves obtained from the BMS telemetry are typically incomplete: measurements often begin only after a certain initial SoC and stop well before reaching full charge. As a result, the original profiles do not span the entire $1\%$--$100\%$ SoC range.

To ensure a consistent charging-power representation across vehicles, each recorded profile is completed at the beginning and end of the SoC domain. Linear extrapolation is applied locally at the session boundaries, preserving the trend of the closest available measurements and avoiding artificial shape assumptions. To avoid unrealistically small values at the boundaries, a minimum power floor is enforced.

Unlike slow or semi-fast AC charging---where the delivered power is approximately constant or follows a simple two-phase (CC--CV) pattern~\cite{DIAZLONDONO2024}---the DC fast and ultra-fast charging profiles studied here exhibit complex behaviors. The charging power is neither constant across the SoC range nor uniformly decreasing after a fixed threshold; instead, each EV’s BMS implements a proprietary control strategy that results in highly diverse power–SoC shapes. Accordingly, completing the profiles over a common SoC domain is required to enable a consistent comparison across EV models and segments.

To facilitate interpretation, the charging profiles are classified according to the maximum accepted DC charging power of each EV. Following the IEA classification, charging infrastructure is differentiated into fast charging (22--150~kW) and ultra-fast charging ($>150$~kW)~\cite{IEA2026}. Since the EV IDs were organized in ascending order of maximum accepted charging power, lower vehicle IDs correspond to EVs with lower charging capability, while higher IDs correspond to EVs capable of operating at higher fast- and ultra-fast charging power levels.

Based on this organization, the profiles are grouped as follows:

\begin{itemize}

    \item \textbf{Group~1:} vehicle IDs 1--42, corresponding to EVs typically served by standard DC fast-charging stations, with:
    \[
    22~\mathrm{kW} < P_{\mathrm{max}} \leq 90~\mathrm{kW}.
    \]

    \item \textbf{Group~2:} vehicle IDs 43--90, representing EVs compatible with high-power DC fast chargers, with:
    \[
    90~\mathrm{kW} < P_{\mathrm{max}} \leq 150~\mathrm{kW}.
    \]

    \item \textbf{Group~3:} vehicle IDs 91--141, corresponding to EVs designed for ultra-fast DC charging infrastructure, with:
    \[
    P_{\mathrm{max}} > 150~\mathrm{kW}.
    \]

\end{itemize}

\begin{figure}
 \centering
 \includegraphics[width=\columnwidth]{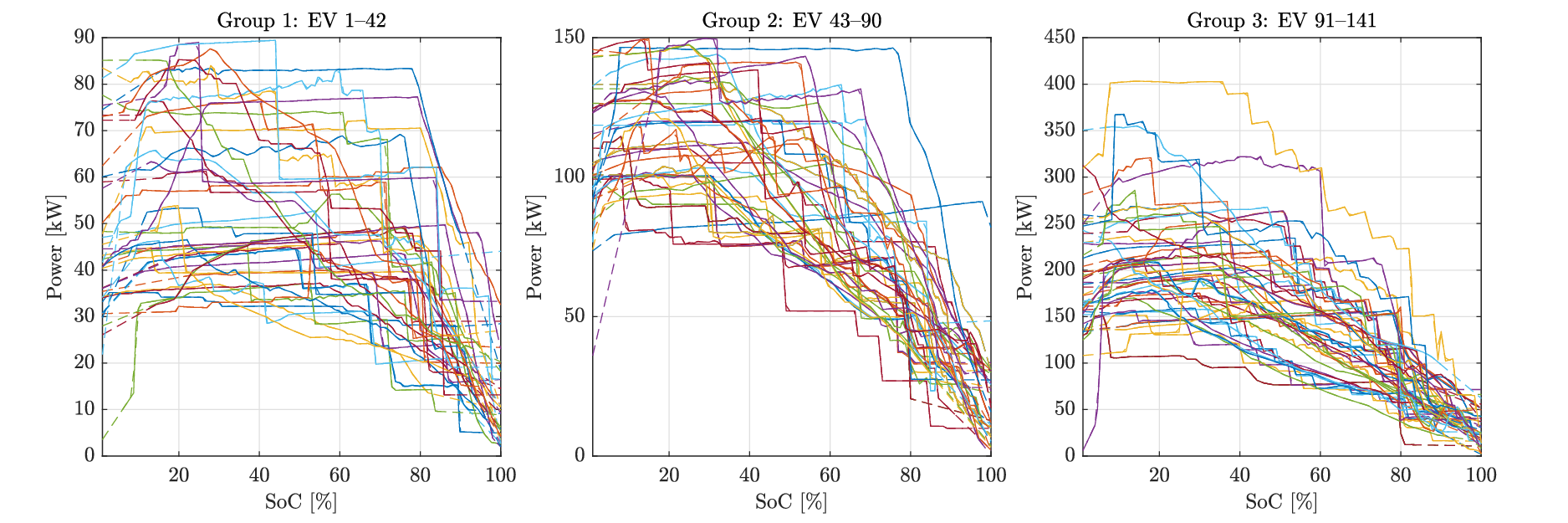}
 \caption{Measured Power--SoC charging curves for the three EV categories used in this study. Dashed segments at the beginning and end indicate extrapolated regions introduced to complete the profiles.}
 \label{fig:EVPowVsSoC}
 \end{figure}

\figurename~\ref{fig:EVPowVsSoC} presents the Power–SoC curves for the three groups, highlighting distinct charging behaviors. Group~1 EVs sustain moderate power levels over extended SoC intervals, Group~2 vehicles achieve higher peaks followed by noticeable power tapering, whereas Group~3 exhibits pronounced initial power bursts prior to BMS-imposed power reduction.

\begin{figure}
 \centering
 \includegraphics[width=\columnwidth]{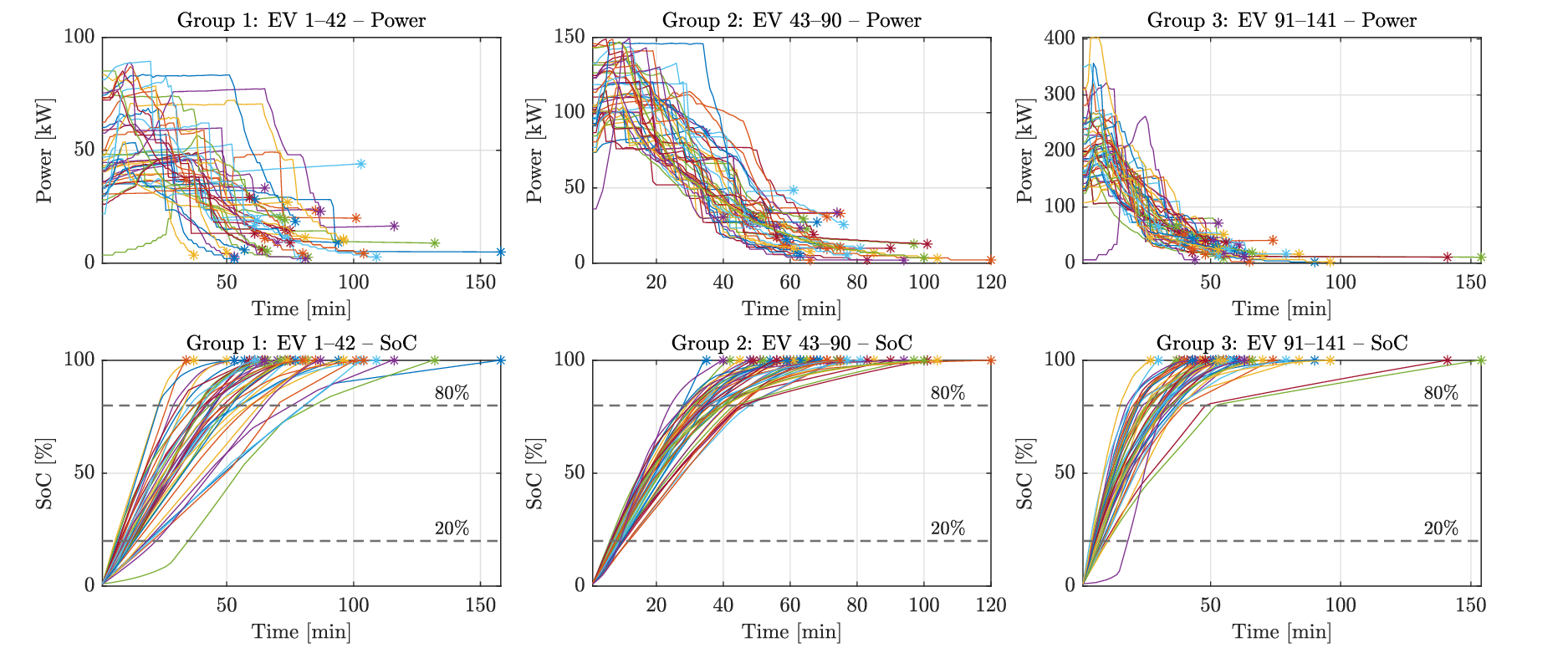}
 \caption{Charging power (top row) and SoC evolution (bottom row) for the three EV groups.}
 \label{fig:Gr_SOCTim}
 \end{figure}

Using the same grouping, \figurename~\ref{fig:Gr_SOCTim} reports the charging profiles in the time domain. The top row shows the charging-power evolution, while the bottom row displays the corresponding SoC trajectories. Across all groups, the power curves confirm the heterogeneous behavior imposed by the various BMS control strategies:

\begin{itemize}
    \item Group 1: smoother power transitions and longer constant-power phases, resulting in gradual SoC increments.
    \item Group 2: higher initial power levels followed by earlier tapering as the SoC increases.
    \item Group 3: very steep power peaks (often above 250--300~kW) and the fastest early SoC growth, followed by strong current limiting near high SoC.
\end{itemize}

The SoC trajectories further clarify these differences. EVs capable of faster charging (Groups~2 and~3) show a rapid initial increase in SoC, reflecting their ability to absorb higher power levels at low SoC. As the SoC increases, all groups progressively transition to slower SoC increments, in line with the expected CC–CV charging regime. Dashed horizontal lines at $20\%$ and $80\%$ SoC delimit the typical fast-charging operating window, emphasizing the region where flexibility and BMS behavior are most relevant.

\subsection{Synthetic EV Discharging Profiles}
\label{subsec:synthetic_discharge}

The quantification of bidirectional flexibility requires a representation of discharge power as a function of SoC. However, the available dataset contains only measured charging Power--SoC curves, while real-world discharging data remain scarce due to the still limited adoption of commercial vehicle-to-grid functionality, currently available in less than 1.5\% of EV models worldwide~\cite{IEA2026}. Since charging and discharging are subject to asymmetric BMS constraints, discharge capability cannot be inferred through a simple sign inversion of the charging profile.

The quantification of bidirectional flexibility requires a representation of discharge power as a function of the SoC. However, the available dataset provides only measured charging Power--SoC curves. Since charging and discharging are subject to asymmetric BMS constraints, the discharge capability cannot be inferred by a simple sign inversion of the charging profile.

Following established voltage-limited battery models and short-horizon approximations based on equivalent circuit frameworks (e.g.,~\cite{Plett2004, Wentao2025, Matthieu2012}), the peak power capability can be interpreted through a simplified voltage-limited representation, in which the terminal voltage is approximated as
\begin{equation}
V_t = OCV(z) - I\,R(z),
\end{equation}
where $z\in[0,1]$ denotes the SoC, $OCV(z)$ is the open-circuit voltage and $R(z)$ is an effective DC resistance. Models of this type are widely used in BMS applications and reflect how terminal voltage and internal resistance vary with SoC, affecting maximum current and power capabilities.

During discharge ($I>0$), the BMS must ensure that the terminal voltage does not drop below a minimum admissible value $V_{\min}$. Enforcing this constraint yields a voltage-limited maximum discharge current:
\begin{equation}
I_{\mathrm{dis,max}}(z)
=
\frac{OCV(z)-V_{\min}}{R_{\mathrm{dis}}(z)},
\end{equation}
and the corresponding discharge power envelope
\begin{equation}
P_{\mathrm{dis,max}}(z)
=
V_{\min}\, I_{\mathrm{dis,max}}(z)
=
\frac{V_{\min}\!\left(OCV(z)-V_{\min}\right)}{R_{\mathrm{dis}}(z)}.
\end{equation}

Since $OCV(z)$ is a monotonically increasing function of $z$ over most of the operating range, the available discharge voltage margin $\bigl(OCV(z)-V_{\min}\bigr)$ is largest at high SoC and progressively shrinks toward low SoC. Consequently, the admissible discharge power naturally decreases as the battery depletes. This behavior is complementary to the measured charging profiles, whose power typically decreases at high SoC as the terminal voltage approaches its upper bound. Experimental peak power test methods further indicate that maximum power capability varies markedly with SoC and operating conditions, reinforcing the need for SoC-dependent characterization.

Based on this physical rationale, synthetic discharge Power--SoC profiles are constructed to be SoC-dependent, monotonic, and consistent with the envelope implied by the above formulation, while preserving the characteristic scaling and overall shape observed in the measured charging data. These profiles are not intended to reproduce proprietary BMS logic, but to provide a physically consistent and conservative representation of discharge capability suitable for flexibility assessment.

For reference, the charging power can be interpreted through the complementary voltage-limited condition imposed by the upper terminal-voltage bound $V_{\max}$. Under this constraint, the maximum admissible charging current magnitude can be written as
\begin{equation}
I_{\mathrm{ch,max}}(z)
=
\frac{V_{\max}-OCV(z)}{R_{\mathrm{ch}}(z)},
\end{equation}
leading to the corresponding charging power envelope
\begin{equation}
P_{\mathrm{ch,max}}(z)
=
V_{\max}\, I_{\mathrm{ch,max}}(z).
\end{equation}


\subsubsection{Synthetic discharge generation from measured charging curves}
\label{subsubsec:synthetic_generation}

Let $P^{\mathrm{ch}}_j(s)$ denote the charging Power--SoC profile of vehicle $j$, defined on the discrete SoC grid $s\in\{1,\dots,100\}$ (in \%). The synthetic discharge profile $P^{\mathrm{dis}}_j(s)$ is generated as
\begin{equation}
P^{\mathrm{dis}}_j(s)
=
\mathcal{R}_{\Delta P_{\max}}\!\left[
\min\!\Bigl(
\beta\,\mathcal{I}\!\left[P^{\mathrm{ch}}_j\right]\! \tilde{s},
\;
Q_{0.9}\!\left(\left\{P^{\mathrm{ch}}_j(s)\right\}_{s=1}^{100}\right)
\Bigr)
\right],
\label{eq:pdis_full}
\end{equation}
where $\mathcal{R}_{\Delta P_{\max}}[\cdot]$ denotes a slope-limiting operator along the SoC axis. The individual components of~\eqref{eq:pdis_full} are described below.
\begin{itemize}
    \item \textit{Mirror mapping.} The discharge SoC index is obtained by reflecting the charging curve around a central SoC value $z_0$,
    \begin{equation}
    \tilde{s} = \Pi_{[1,100]}(2z_0-s),
    \end{equation}
    where $\Pi_{[1,100]}(\cdot)$ denotes projection (clipping) onto the admissible SoC domain. In this work, $ z_0 = 50\%$, which yields a symmetric mapping.

    \item \textit{Interpolation and amplitude scaling.} The operator $\mathcal{I}[\cdot]$ denotes linear interpolation on the discrete SoC grid, ensuring a continuous mapping between integer SoC levels. A global scaling factor $ \beta < 1$ is applied to represent conservative sustained-discharge capability, while preserving the relative shape of the charging profile across vehicles.

    \item \textit{Soft peak-power cap.} To prevent isolated charging peaks from being mirrored into unrealistically large discharge values, a per-vehicle soft cap is enforced using the 90th percentile of the charging profile,
    \begin{equation}
    Q_{0.9}\!\left(\left\{P^{\mathrm{ch}}_j(s)\right\}_{s=1}^{100}\right),
    \end{equation}
    which bounds the discharge power without imposing a hard saturation.

    \item \textit{Ramp-rate limitation in the SoC domain.} The operator $\mathcal{R}_{\Delta P_{\max}}[\cdot]$ enforces a maximum variation $\Delta P_{\max}$ (in kW) per 1\% SoC step, limiting abrupt SoC-to-SoC power changes that are unlikely under BMS-controlled transitions. This operation ensures smooth and physically plausible discharge trajectories.
\end{itemize}

\begin{figure}
 \centering
 \includegraphics[width=\columnwidth]{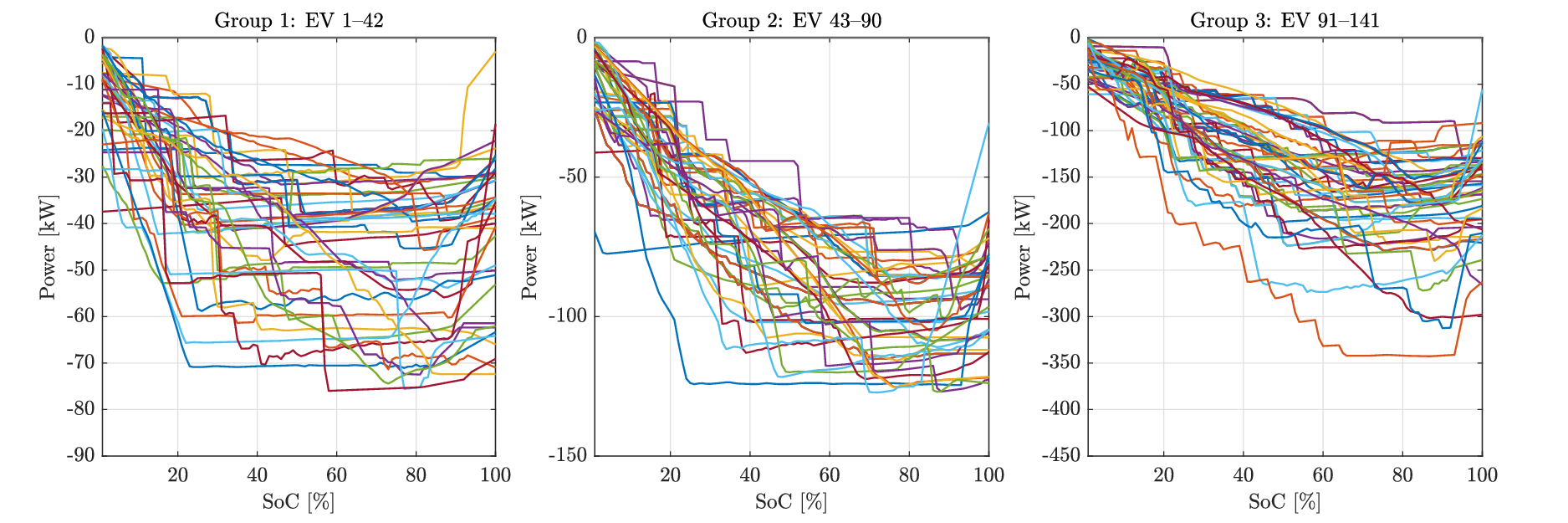}
 \caption{Synthetic discharge Power--SoC curves for the three EV groups.}
 \label{fig:EVPowVsSoC_dis}
 \end{figure}

Using the same EV grouping adopted for the charging analysis, \figurename~\ref{fig:EVPowVsSoC_dis} reports the synthetic discharge Power--SoC curves for the three groups. Compared with the charging case, the discharge profiles exhibit an opposite SoC dependence, with higher power levels sustained at high SoC and a progressive reduction as the battery approaches low SoC. Although synthetically generated, the curves retain a clear degree of inter-vehicle variability, capturing differences in charging characteristics as well as the effect of the applied shaping constraints.

\begin{figure}
 \centering
 \includegraphics[width=\columnwidth]{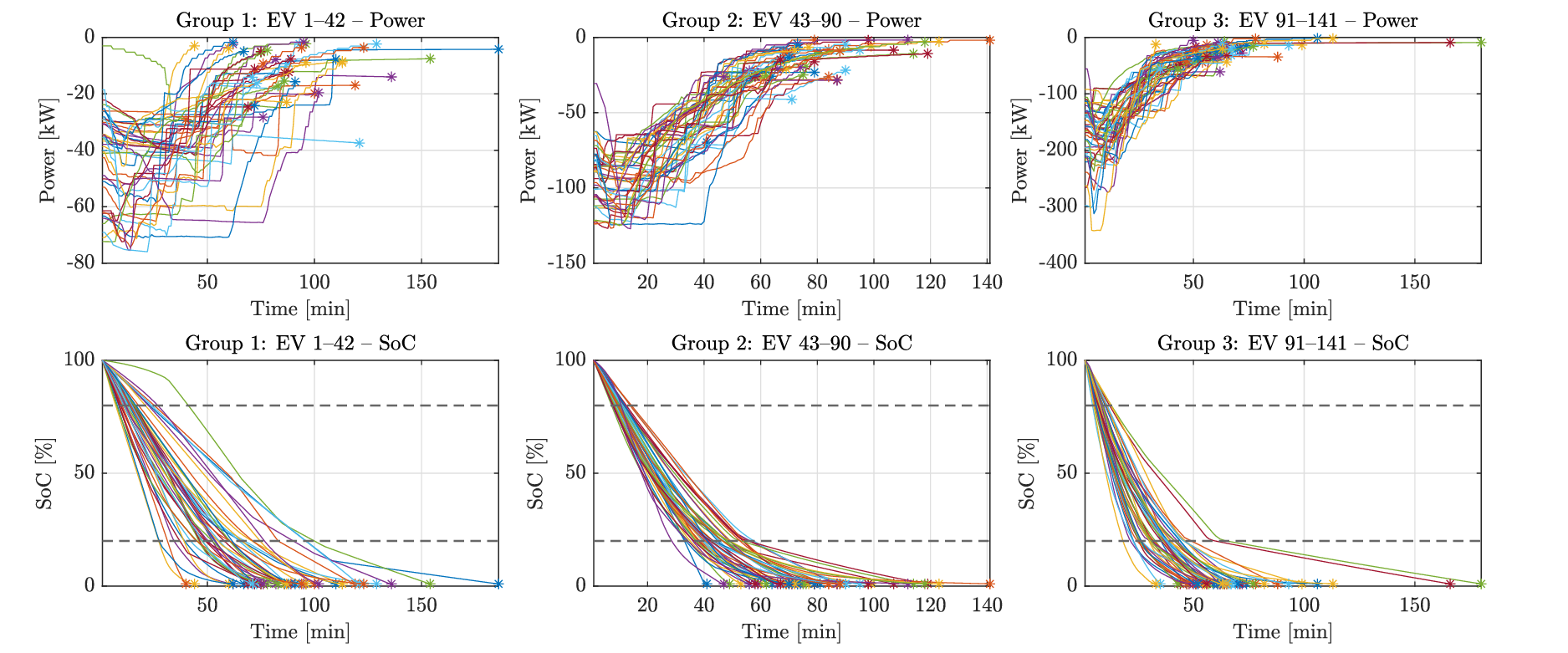}
 \caption{Synthetic discharge trajectories in the time domain for the three EV groups. Top: discharge power evolution. Bottom: corresponding SoC trajectories.}
 \label{fig:Gr_SOCTim_dis}
 \end{figure}

\figurename~\ref{fig:Gr_SOCTim_dis} illustrates the synthetic discharge trajectories in the time domain for the three EV groups. The discharge power evolution is shown in the upper panels, while the corresponding SoC trajectories are reported in the lower panels. The results reveal different discharge dynamics across vehicle categories.

These time-domain patterns are consistent with the Power--SoC characteristics observed in the synthetic discharge envelopes and provide additional insight into how bidirectional flexibility manifests over the connection horizon for different EV categories.

\subsubsection{Assumptions and limitations}

The synthetic discharge profiles are obtained from the charging Power–SoC curves using the mapping and shaping rule defined in \eqref{eq:pdis_full}. This construction is grounded in the well-known asymmetry between charging and discharging capabilities imposed by terminal-voltage margins. At the same time, it does not explicitly compute discharge power from internal battery variables—such as terminal voltage, DC resistance, or temperature—which are not available in the adopted dataset.

As a result, the derived discharge curves should be interpreted as conservative and physically plausible envelopes for system-level bidirectional flexibility assessment, rather than as manufacturer-accurate reproductions of proprietary BMS behavior. In particular, thermal transients, aging or state-of-health effects, and detailed current-limiting logic are not modeled, and localized features observed in measured charging profiles (e.g., short plateaus or step-like events) may not have a direct counterpart during discharge.

\section{Flexibility of Fast and Ultra-Fast EV Chargers}
\label{sec:flexibility}

This section defines and quantifies the flexibility potential of fast and ultra-fast DC charging sessions. Flexibility is understood as the ability of an EV to temporally reshape its power exchange with the grid while satisfying user-defined SoC targets and respecting battery-protection constraints. In the context of slow AC charging, flexibility has been traditionally characterized through upward and downward reserves enabled by direct modulation of the charging power \cite{DiazLondono2025}. However, unlike slow AC charging—where power setpoints can often be freely adjusted by the charging infrastructure—DC fast charging is predominantly governed by the vehicle-side battery BMS. Consequently, flexibility cannot be expressed as arbitrary power modulation, but rather as a constrained energy reshaping capability determined by the Power--SoC characteristics of each EV and the available connection time.

\begin{figure}
     \centering
     \includegraphics[width=\linewidth]{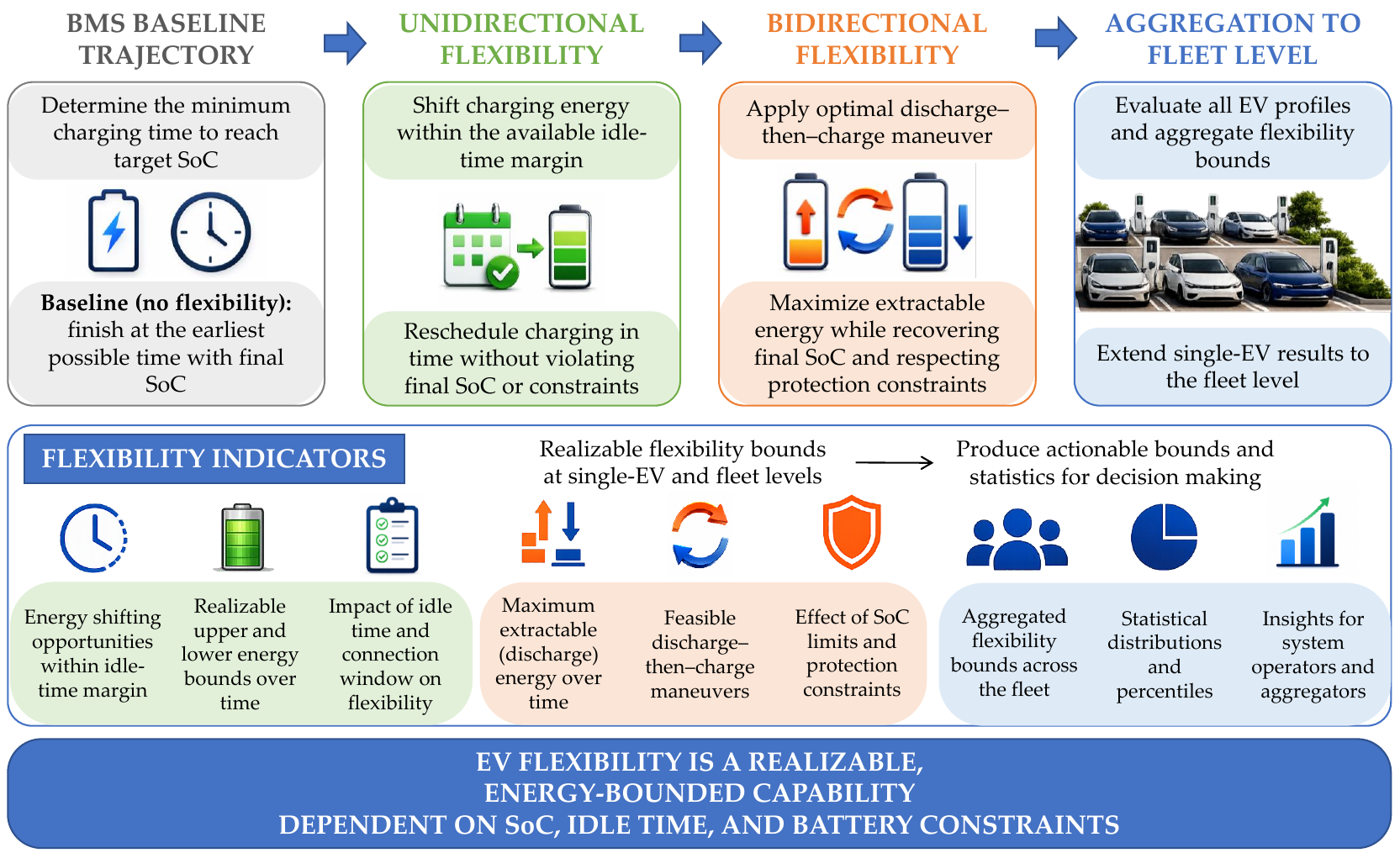}
     \caption{Framework for quantifying flexibility in fast and ultra-fast DC charging under BMS-governed Power--SoC dynamics.}
     \label{fig:FlexFramework}
 \end{figure}

\figurename~\ref{fig:FlexFramework} summarizes the step-by-step evaluation logic adopted in this section. First, the BMS-imposed baseline trajectory defines the minimum charging time to reach the target SoC. If additional connection slack is available, (i) unidirectional flexibility is quantified by shifting charging energy within the idle-time margin, and (ii) bidirectional flexibility is quantified by the optimal discharge--then--charge maneuver that maximizes the extractable discharge energy while ensuring recovery to the final SoC and enforcing conservative battery-protection constraints. The same procedure is computed at the single-EV level and then aggregated across EVs to obtain fleet-level flexibility maps and statistics.

Throughout this section, the charging trajectory of vehicle $j$ is represented in discrete time as
\[
\mathrm{SoC}_j(k), \qquad P^{\mathrm{ch}}_j(k),
\]
where $k$ denotes the time index (with one-minute resolution), $\mathrm{SoC}_j(k)\in[1,100]$ is the battery SoC, and $P^{\mathrm{ch}}_j(k)$ is the charging power. User requirements are defined by an initial $\mathrm{SoC}_{\mathrm{ini}}$ and final $\mathrm{SoC}_{\mathrm{fin}}$ state of charge.

The effective start and end indices of the mandatory charging segment are identified from the trajectories presented in the previous section as
\begin{equation}
k_{\mathrm{ini}} = \arg\min_k \left| \mathrm{SoC}_j(k) -
\mathrm{SoC}_{\mathrm{ini}} \right|,
\quad
k_{\mathrm{fin}} = \arg\min_k \left| \mathrm{SoC}_j(k) -
\mathrm{SoC}_{\mathrm{fin}} \right|.
\end{equation}

The minimum time required to satisfy the charging request is then
\begin{equation}
T_j = k_{\mathrm{fin}} - k_{\mathrm{ini}},
\end{equation}
and the corresponding delivered energy is
\begin{equation}
E_j =
\sum_{k=k_{\mathrm{ini}}+1}^{k_{\mathrm{fin}}}
\frac{P^{\mathrm{ch}}_j(k)}{\gamma},
\label{eq:energy_interval}
\end{equation}
where $\gamma=60$ converts power samples into energy for the adopted one-minute resolution. Flexibility can only arise if the total connection time exceeds $T_j$, thereby creating an idle-time margin that can be exploited for energy reshaping.


\subsection{Unidirectional Charging Flexibility Potential}
\label{subsec:unidir_flex}

This subsection focuses on quantifying the unidirectional flexibility potential of fast and ultra-fast DC charging sessions. Flexibility is interpreted here as an energy-based quantity, representing the amount of charging energy that can be temporally shifted within the connection window while still satisfying the user-defined final SoC requirement. Rather than defining instantaneous power reserves, the analysis therefore aims to identify the bounds of the energy that can be shifted over time.

By definition, the minimum shiftable energy is zero, as charging can always proceed without temporal adjustment. In contrast, the maximum shiftable energy is time-dependent, since it is constrained by both the remaining connection duration and the admissible charging trajectory. As a result, flexibility at any given time can be described as the range between two limiting envelopes: an upper bound on the maximum achievable flexibility and a lower bound on this maximum.

It is first noted that no flexibility exists strictly within the mandatory charging interval $[k_{\mathrm{ini}},k_{\mathrm{fin}}]$, since all power delivered in this period is required to reach the target $\mathrm{SoC}_{\mathrm{fin}}$. Unidirectional flexibility arises only when the EV remains connected beyond this minimum charging time, creating an additional idle-time margin $\Delta t$ that can be exploited to reshape the charging profile.

Within this extended horizon, the flexibility potential is quantified by analyzing how much energy can be reallocated within any $\Delta t$-long segment of the mandatory charging trajectory. To this end, a sliding-window energy analysis is applied to the charging power profile $P^{\mathrm{ch}}_j(k)$. For a given window length $\Delta t$, the sliding-window energy profile of vehicle $j$ is defined as
\begin{equation}
S_j^{(\Delta t)}(k) =
\sum_{m=0}^{\Delta t-1} P^{\mathrm{ch}}_j(k+m),
\qquad
k = k_{\mathrm{ini}}+1,\ldots,k_{\mathrm{fin}}-\Delta t+1,
\label{eq:sliding_energy}
\end{equation}
which represents the total energy absorbed over a $\Delta t$-minute block starting at time index $k$.

The $\Delta t$-windows associated with the upper and lower bounds of the maximum shiftable energy are identified as
\begin{equation}
k^{\mathrm{UB}} = \arg\max_k S_j^{(\Delta t)}(k),
\qquad
k^{\mathrm{LB}} = \arg\min_k S_j^{(\Delta t)}(k),
\end{equation}
from which the corresponding energy bounds are obtained:
\begin{equation}
E_{j,\mathrm{uni}}^{\mathrm{UB}}(\Delta t) =
\frac{1}{\gamma} S_j^{(\Delta t)}(k^{\mathrm{UB}}),
\qquad
E_{j,\mathrm{uni}}^{\mathrm{LB}}(\Delta t) =
\frac{1}{\gamma} S_j^{(\Delta t)}(k^{\mathrm{LB}}),
\end{equation}
where $k^{\mathrm{UB}}$ and $k^{\mathrm{LB}}$ identify the time windows associated with the upper and lower bounds of the \emph{maximum shiftable energy}, and $\gamma = 60$ corresponds to the adopted one-minute time resolution. Note that the minimum flexibility is zero by definition; therefore, $k^{\mathrm{LB}}$ does not represent a minimum flexibility value, but the lower bound of the maximum achievable energy shift within the considered time window.

To allow comparison across vehicles with different battery capacities, these quantities are normalized as
\begin{equation}
F_{j,\mathrm{uni}}^{\mathrm{UB}}(\Delta t) =
\frac{E_{j,\mathrm{uni}}^{\mathrm{UB}}(\Delta t)}{C_j},
\qquad
F_{j,\mathrm{uni}}^{\mathrm{LB}}(\Delta t) =
\frac{E_{j,\mathrm{uni}}^{\mathrm{LB}}(\Delta t)}{C_j},
\end{equation}
where $C_j$ denotes the battery capacity of vehicle $j$.

The metrics $F_{j,\mathrm{uni}}^{\mathrm{UB}}$ and $F_{j,\mathrm{uni}}^{\mathrm{LB}}$ define the upper and lower bounds of the \emph{maximum achievable SoC variation} within a $\Delta t$-long operational window. They therefore characterize the feasible range of unidirectional flexibility in normalized (SoC) terms.

\begin{figure}
     \centering
     \includegraphics[width=\columnwidth]{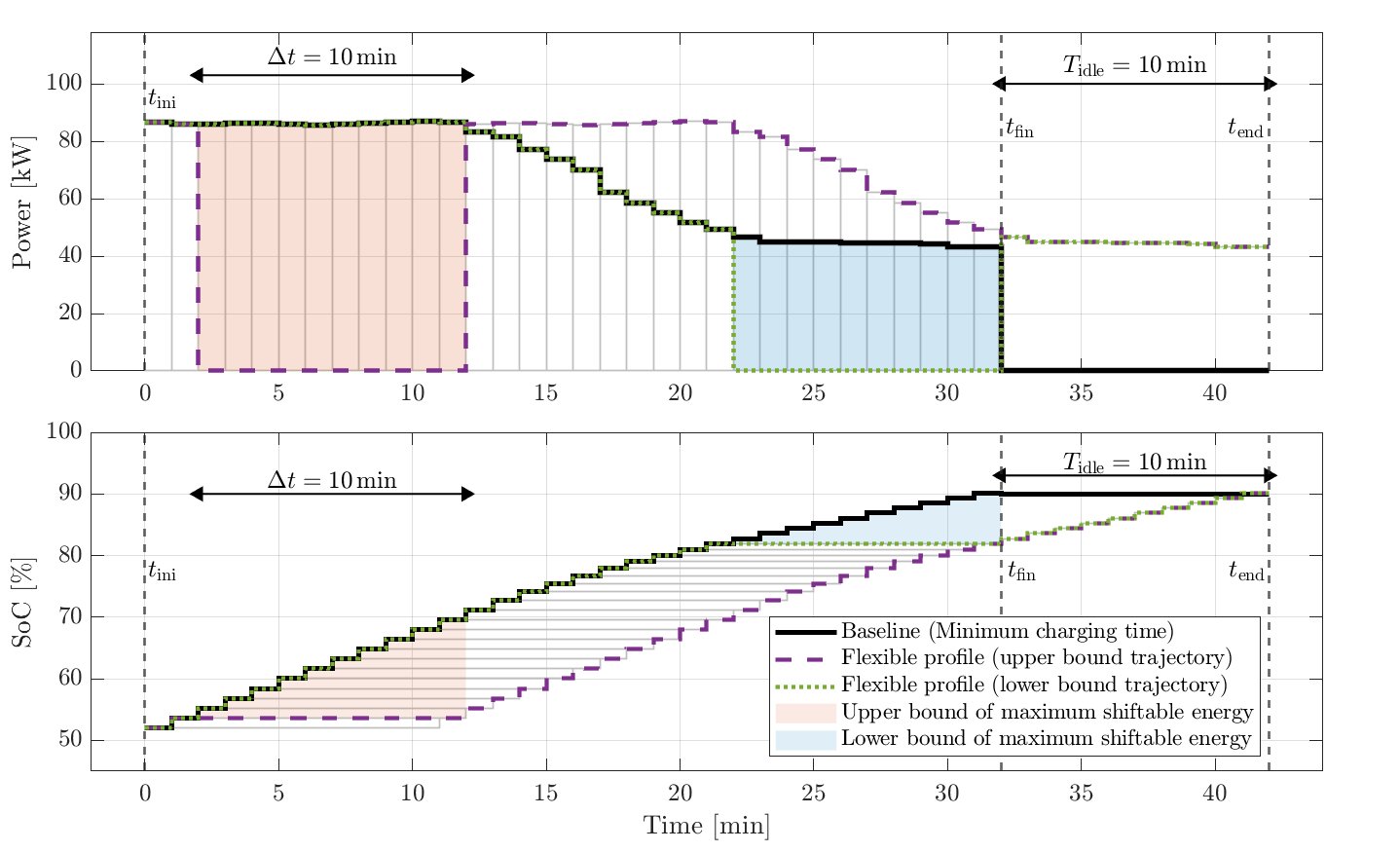}
     \caption{Conceptual illustration of the sliding-window method used to quantify unidirectional flexibility as an energy-bounded process, defined by the upper and lower bounds of the maximum shiftable energy, $E_{j,\mathrm{uni}}^{\mathrm{UB}}$ and $E_{j,\mathrm{uni}}^{\mathrm{LB}}$.}
     \label{fig:FlexConceptUni}
 \end{figure}

\figurename~\ref{fig:FlexConceptUni} presents the unidirectional flexibility quantification for a representative EV charging session as a conceptual illustration. In this example, the vehicle connects at an initial SoC of approximately $52\%$ and reaches the target $\mathrm{SoC}_{\mathrm{fin}}=90\%$ after a mandatory charging time of $T_j = 32$~min. The EV remains connected for an additional idle period of $T_{\mathrm{idle}}=\Delta t=10$~min, which enables flexibility. The shaded regions correspond to the $\Delta t$-long windows associated with the upper (orange) and lower (blue) bounds of the maximum energy content along the mandatory charging trajectory, yielding $E_{j,\mathrm{uni}}^{\mathrm{UB}}=14.4$~kWh and $E_{j,\mathrm{uni}}^{\mathrm{LB}}=7.4$~kWh, respectively. These values define the bounds of the maximum shiftable charging energy within the considered time window. The corresponding flexible profiles illustrate how this energy can be temporally redistributed within the idle-time margin while still satisfying the final SoC requirement.

Overall, the example highlights that unidirectional flexibility is an energy-bounded quantity jointly determined by the available idle time and the heterogeneity of the charging power profile. All feasible charging trajectories are necessarily confined within these bounds. Within this region, a continuum of feasible charging profiles exists, characterized by different temporal redistributions of energy and intermediate power levels, as long as the charging trajectories requested by the operator remain within the feasible bounds and the final SoC requirement is satisfied.


\subsection{Bidirectional Flexibility Limits with Idle-Time and Battery-Protection Constraints}
\label{subsec:flex_limits}

While the unidirectional analysis characterizes how charging energy can be temporally reshaped within the idle-time margin, bidirectional flexibility further extends this concept by allowing controlled energy injection back to the grid. In fast and ultra-fast DC charging, such bidirectional operation is strongly constrained by the EV BMS, the available idle time, and explicit battery-protection limits. This subsection formalizes the admissible bidirectional flexibility envelope under these coupled constraints.

\subsubsection{Baseline trajectory and available idle time}

For a given charging session, vehicle $j$ arrives with an initial $\mathrm{SoC}_{\mathrm{ini}}$ and must depart with a target $\mathrm{SoC}_{\mathrm{fin}}$, following the BMS-imposed charging trajectory described in Section~\ref{sec:EV_analysis}. The baseline trajectory corresponds to the mandatory charging process from $\mathrm{SoC}_{\mathrm{ini}}$ to $\mathrm{SoC}_{\mathrm{fin}}$.

Throughout this subsection, time is expressed in continuous form for conceptual clarity, consistently with the illustrative figures. This notation is equivalent to the discrete-time representation adopted earlier, with sampling interval $t_s=1$~min, such that $t = k\,t_s$.

Let $t_{\mathrm{fin},j}$ denote the time instant at which the baseline trajectory reaches $\mathrm{SoC}_{\mathrm{fin}}$ (i.e., $t_{\mathrm{fin},j} = k_{\mathrm{fin},j}\,t_s$). If the EV remains connected beyond this instant, an idle-time margin becomes available. The total connection horizon is thus defined as
\begin{equation}
T_{\mathrm{end},j} = t_{\mathrm{fin},j} + T_{\mathrm{idle}},
\label{eq:Tend}
\end{equation}
where $T_{\mathrm{idle}}$ represents the idle time.

For any intermediate $\mathrm{SoC}_{j,k} \in [\mathrm{SoC}_{\mathrm{ini}}, \mathrm{SoC}_{\mathrm{fin}}]$, reached at time $t_{j,k} = k\,t_s$ along the baseline trajectory, the remaining connection time available for flexibility actions is
\begin{equation}
T_{\mathrm{rem},j}(\mathrm{SoC}_{j,k}) =
T_{\mathrm{end},j} - t_{j,k}.
\label{eq:Trem}
\end{equation}
This quantity represents the effective time budget within which any bidirectional maneuver must be fully executed.

\subsubsection{Candidate maneuver and feasibility condition}

Bidirectional flexibility is represented through a single \emph{discharge--then--charge} maneuver initiated at $\mathrm{SoC}_{j,k}$ along the baseline trajectory. Starting from $\mathrm{SoC}_{j,k}$, the EV discharges down to a minimum $\mathrm{SoC}_{j,\min}$ and is then recharged back to $\mathrm{SoC}_{\mathrm{fin}}$, such that the final SoC target is met exactly at the disconnection time $T_{\mathrm{end},j}$. This maneuver is introduced as a representative construction, assuming that any admissible combination of charging and discharging actions remains bounded by the same time and energy constraints. As such, the formulation captures the feasible flexibility limits without restricting the analysis to a specific trajectory.

Let $T_{\mathrm{dis},j}(\mathrm{SoC}_{j,k},\mathrm{SoC}_{j,\min})$ and $T_{\mathrm{ch},j}(\mathrm{SoC}_{j,\min},\mathrm{SoC}_{\mathrm{fin}})$ denote the time required for the discharge and recharge phases, respectively, as determined by the Power--SoC trajectories. The maneuver is feasible if
\begin{equation}
T_{\mathrm{dis},j}(\mathrm{SoC}_{j,k},\mathrm{SoC}_{j,\min})
+
T_{\mathrm{ch},j}(\mathrm{SoC}_{j,\min},\mathrm{SoC}_{\mathrm{fin}})
\;\le\;
T_{\mathrm{rem},j}(\mathrm{SoC}_{j,k}).
\label{eq:feasibility_time}
\end{equation}

This condition highlights that bidirectional flexibility is inherently time-limited: deeper discharge levels can only be exploited if sufficient idle time is available to subsequently recover the required SoC before departure.

\subsubsection{Battery-protection constraints}

In addition to time feasibility, battery-protection constraints are enforced to avoid excessive degradation during bidirectional operation.
\begin{itemize}
    \item \textit{Lower SoC bound.} The minimum SoC reached during the maneuver must satisfy
    \begin{equation}
    \mathrm{SoC}_{j,\min} \ge \mathrm{SoC}_{\mathrm{prot}},
    \label{eq:soc_protection}
    \end{equation}
    where $\mathrm{SoC}_{\mathrm{prot}}$ denotes a conservative lower bound imposed by the BMS or by operational policies.

    \item \textit{Maximum depth-of-discharge constraint.} To further limit battery stress, the total discharge depth within a single charging session is constrained by a maximum admissible depth-of-discharge parameter $\Delta\mathrm{SoC}_{\max}$:
    \begin{equation}
    \mathrm{SoC}_{j,k} - \mathrm{SoC}_{j,\min} \le \Delta\mathrm{SoC}_{\max}.
    \label{eq:dod_limit}
    \end{equation}
    This constraint is imposed at the session level to avoid more than one equivalent full charge--discharge cycle per day, acknowledging that a non-negligible fraction of the battery throughput is inherently consumed by vehicle operation between charging events. The parameter $\Delta\mathrm{SoC}_{\max}$ represents the maximum fraction of the battery capacity that can be allocated to grid-oriented services during a single connection, with the remaining capacity reserved for mobility-related usage.

    For illustration, and consistently with conservative assumptions commonly adopted in degradation-aware flexibility studies~\cite{Fernandes2025}, this work considers $\Delta\mathrm{SoC}_{\max}=60\%$. Under this assumption, if a maneuver is initiated at $\mathrm{SoC}_k = 90\%$, the minimum admissible state of charge is $\mathrm{SoC}_{\min} = 30\%$, even if longer idle times would otherwise allow deeper discharge.
\end{itemize}

The admissible value of $\mathrm{SoC}_{j,\min}$ is therefore determined by the joint satisfaction of the time-feasibility condition \eqref{eq:feasibility_time} and the battery-protection constraints \eqref{eq:soc_protection}--\eqref{eq:dod_limit}.

\subsubsection{Energy-based flexibility limits}

For each feasible starting point $\mathrm{SoC}_{j,k}$, the discharge and recharge energies associated with the bidirectional maneuver are computed using the real charging and synthetic discharge profiles:
\begin{equation}
E_{j,\mathrm{dis}}(\mathrm{SoC}_{j,k}) =
\sum_{n \in \mathcal{D}(\mathrm{SoC}_{j,k},\mathrm{SoC}_{j,\min})}
P^{\mathrm{dis}}_j(n)\, t_{\mathrm{s}},
\end{equation}
\begin{equation}
E_{j,\mathrm{ch}}(\mathrm{SoC}_{j,k}) =
\sum_{n \in \mathcal{C}(\mathrm{SoC}_{j,\min},\mathrm{SoC}_{\mathrm{fin}})}
P^{\mathrm{ch}}_j(n)\, t_{\mathrm{s}},
\end{equation}
where $\mathcal{D}(\cdot)$ and $\mathcal{C}(\cdot)$ denote the discrete-time indices corresponding to the discharge and charge segments, respectively, and $t_{\mathrm{s}}$ is the sampling interval.

By sweeping $\mathrm{SoC}_{j,k}$ over the interval $[\mathrm{SoC}_{\mathrm{ini}}, \mathrm{SoC}_{\mathrm{fin}}]$, the admissible range of the maximum bidirectional flexibility is obtained. The corresponding bounds are defined as
\begin{equation}
E_{j,\mathrm{bi}}^{\mathrm{UB}}
=
\max_{\mathrm{SoC}_{j,k}\in[\mathrm{SoC}_{\mathrm{ini}},\,\mathrm{SoC}_{\mathrm{fin}}]}
E_{j,\mathrm{dis}}(\mathrm{SoC}_{j,k}),
\end{equation}
\begin{equation}
E_{j,\mathrm{bi}}^{\mathrm{LB}}
=
\min_{\mathrm{SoC}_{j,k}\in[\mathrm{SoC}_{\mathrm{ini}},\,\mathrm{SoC}_{\mathrm{fin}}]}
E_{j,\mathrm{dis}}(\mathrm{SoC}_{j,k}),
\end{equation}
where $E_{j,\mathrm{bi}}^{\mathrm{UB}}$ corresponds to the upper bound of the maximum extractable discharge energy, associated with the deepest feasible discharge compatible with idle-time and battery-protection constraints, while $E_{j,\mathrm{bi}}^{\mathrm{LB}}$ represents the lower bound of the maximum extractable discharge energy within the admissible operating region. Note that the minimum flexibility is zero by definition; therefore, these bounds characterize the range of the maximum achievable bidirectional flexibility rather than minimum and maximum flexibility values.

This formulation makes explicit how bidirectional flexibility in fast and ultra-fast DC charging emerges as an energy-bounded quantity jointly determined by the underlying Power--SoC characteristics, available idle time, and conservative battery-protection limits.

\begin{figure}
     \centering
     \includegraphics[width=\columnwidth]{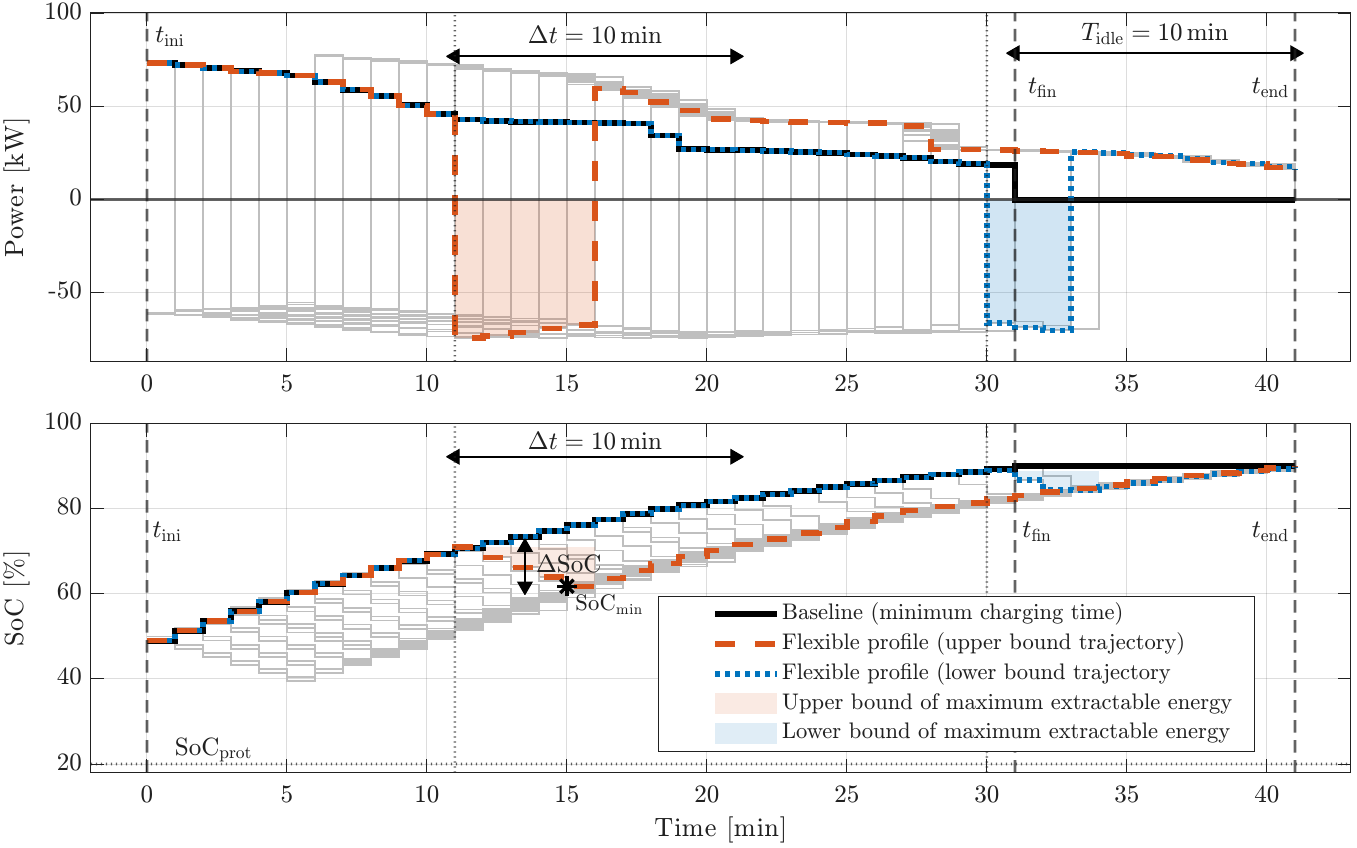}
     \caption{Conceptual illustration of bidirectional flexibility for a representative fast-charging session. Top: power trajectories. Bottom: SoC evolution. Shaded areas indicate the discharged energy associated with the upper and lower bounds of the maximum extractable energy within the idle-time margin $\Delta t$.}
     \label{fig:FlexConceptBid}
 \end{figure}

\figurename~\ref{fig:FlexConceptBid} illustrates the bidirectional flexibility quantification for a representative EV fast-charging session, explicitly accounting for idle-time availability and battery-protection constraints. The EV reaches the target $\mathrm{SoC}_{\mathrm{fin}}=90\%$ following the mandatory charging trajectory and remains connected for an additional idle period, which enables bidirectional flexibility within the feasible bounds.

The black curves represent the baseline trajectory, consisting of charging up to $\mathrm{SoC}_{\mathrm{fin}}$ followed by idle connection. The dashed colored curves illustrate two feasible bidirectional maneuvers initiated at different intermediate states $\mathrm{SoC}_{j,k}$ along the baseline. The shaded regions highlight the energy exchanged during the discharge phase, corresponding to the upper (orange) and lower (blue) bounds of the maximum extractable energy.

In the upper-bound case, the maneuver is initiated at $\mathrm{SoC}_{j,k}=64\%$, leading to a discharge of $T_{\mathrm{dis},j}=4.64$~min with a mean power of $-64.61$~kW and a discharged energy of $E_{j,\mathrm{bi}}^{\mathrm{UB}}=6.46$~kWh. In contrast, the lower-bound case starts at $\mathrm{SoC}_{j,k}=88\%$, resulting in a shorter discharge of $T_{\mathrm{dis},j}=1.36$~min at a higher mean power of $-84.36$~kW and a discharged energy of $E_{j,\mathrm{bi}}^{\mathrm{LB}}=2.81$~kWh.

The SoC panel further highlights the active constraints: the lower bound $\mathrm{SoC}_{\mathrm{prot}}$ limits the depth of discharge, while the vertical arrow $\Delta\mathrm{SoC}_{\max}$ and the marker $\mathrm{SoC}_{\min}$ indicate the maximum admissible SoC excursion reached in the upper-bound case. 

Overall, the example shows that bidirectional flexibility is an energy-bounded and time-coupled quantity, strongly dependent on the initiation point $\mathrm{SoC}_{j,k}$ and on the asymmetric discharge and recharge dynamics. All feasible bidirectional trajectories are necessarily confined within these bounds, while a continuum of admissible charging and discharging profiles can be realized within this region.


\subsubsection{Idle-time–dependent flexibility domain and feasibility}
\label{subsec:idle_sensitivity}

The flexibility limits introduced previously are inherently conditioned by the amount of idle time available after the target SoC is reached. To make this dependence explicit, the bidirectional flexibility domain is parametrized with respect to the idle time $T_{\mathrm{idle}}$. Accordingly, the bidirectional maneuver is formulated for a generic idle time $T_{\mathrm{idle}} \in \mathcal{T}$, while keeping $\mathrm{SoC}_{\mathrm{ini}}$ and $\mathrm{SoC}_{\mathrm{fin}}$ fixed. For each $T_{\mathrm{idle}}$ and each intermediate point $\mathrm{SoC}_{j,k}\in[\mathrm{SoC}_{\mathrm{ini}},\mathrm{SoC}_{\mathrm{fin}}]$, the remaining time available to reshape the profile is defined as
\begin{equation}
T_{\mathrm{rem},j}(\mathrm{SoC}_{j,k};T_{\mathrm{idle}})
=
\bigl(t_{\mathrm{fin},j}+T_{\mathrm{idle}}\bigr)-t_{j,k},
\end{equation}
which makes explicit that the flexibility window increases monotonically with $T_{\mathrm{idle}}$ and shrinks as $\mathrm{SoC}_{j,k}$ approaches $\mathrm{SoC}_{\mathrm{fin}}$.

For each pair $(\mathrm{SoC}_{j,k}, T_{\mathrm{idle}})$, the maneuver searches the deepest admissible discharge level $\mathrm{SoC}_{j,\min}^{*}$ that simultaneously satisfies (i) time feasibility and (ii) battery-protection constraints. The following constrained optimization problem formalizes this:
\begin{align}
\mathrm{SoC}_{j,\min}^{*}(\mathrm{SoC}_{j,k}, T_{\mathrm{idle}})
=
\arg\min_{\mathrm{SoC}_{j,\min}} \quad
& \mathrm{SoC}_{j,\min} \\
\text{s.t.} \quad
& \eqref{eq:feasibility_time},\ \eqref{eq:soc_protection},\ \eqref{eq:dod_limit}.
\label{eq:socmin_star}
\end{align}
If no $\mathrm{SoC}_{j,\min}$ satisfies the constraints, the pair $(\mathrm{SoC}_{j,k}, T_{\mathrm{idle}})$ is declared infeasible.

Once the optimal discharge depth $\mathrm{SoC}_{j,\min}^{*}$ is determined, the corresponding maneuver is fully specified. The associated extractable discharge energy is then computed in discrete time as
\begin{equation}
E_{j,\mathrm{dis}}(\mathrm{SoC}_{j,k}, T_{\mathrm{idle}})
=
\sum_{n=0}^{N_{\mathrm{dis}}-1} P^{\mathrm{dis}}_j(n)\, t_{\mathrm{s}},
\label{eq:Edis_map}
\end{equation}
where $\{P^{\mathrm{dis}}_j(n)\}_{n=0}^{N_{\mathrm{dis}}-1}$ denotes the discharge-power segment of the maneuver and $t_{\mathrm{s}}$ is the sampling interval.


\subsubsection{Fleet-level evaluation framework under idle-time constraints}
\label{subsec:fleet_eval_framework}

To extend the bidirectional flexibility formulation from single charging sessions to a fleet-level characterization, the bidirectional maneuver is evaluated for all EV profiles over a discrete set of idle-time values $T_{\mathrm{idle}}\in\mathcal{T}$. This establishes a unified evaluation framework in which flexibility is jointly determined by the EV-specific power--SoC characteristics and the available connection slack beyond the baseline charging requirement.

For each EV $j$, the baseline charging trajectory defines the SoC interval over which bidirectional actions can be initiated. The analysis is carried out over all admissible intermediate operating points $\mathrm{SoC}_{j,k}$ within this interval, using the baseline trajectory to associate each $\mathrm{SoC}_{j,k}$ with its corresponding time stamp $t_{j,k}$ and remaining connection time $T_{\mathrm{rem},j}$.

For a given idle time $T_{\mathrm{idle}}$, a bidirectional maneuver is feasible at $(j,\mathrm{SoC}_{j,k})$ if the time-feasibility condition and the battery-protection constraints are simultaneously satisfied.

Whenever a feasible maneuver exists, the corresponding extractable discharge energy $E_{j,\mathrm{dis}}(\mathrm{SoC}_{j,k},T_{\mathrm{idle}})$ is computed. This quantity constitutes the fundamental energy-based indicator of bidirectional flexibility at a given operating point and idle time.

This framework enables a consistent aggregation of feasibility and extractable energy across SoC operating points, idle-time values, and EVs, forming the basis for the fleet-level bounds, feasibility maps, and performance indicators presented in the following section.

\subsection{Flexibility metrics and indicators}
\label{subsec:flex_kpis}

To enable a consistent and operational quantification of flexibility in fast and ultra-fast EV charging sessions, this study proposes a unified set of indicators defined at the level of each EV and connection scenario. These indicators capture not only the amount of energy that can be shifted or extracted, but also the temporal availability, depth, and feasibility of such actions. As such, they are directly relevant for system operators, aggregators, and energy-management applications seeking to exploit EV flexibility under realistic operational and battery-protection constraints.

The proposed flexibility indicators are summarized as follows.

\paragraph{Time-based availability indicators}
\begin{itemize}
    \item \emph{Minimum charging time} $T_j$, defined as the time required to reach the target state of charge $\mathrm{SoC}_{\mathrm{fin}}$ along the baseline trajectory. This quantity represents the non-flexible portion of the session.
    \item \emph{Available idle time} $T_{\mathrm{idle}}$, corresponding to the connection slack beyond $T_j$. This is a key flexibility indicator, as it defines the time window within which any reshaping or bidirectional action can be accommodated without violating user requirements.
\end{itemize}

\paragraph{Energy-based flexibility indicators}
\begin{itemize}
    \item \emph{Unidirectional energy bounds} $E_{j,\mathrm{uni}}^{\mathrm{LB}}(\Delta t)$ and $E_{j,\mathrm{uni}}^{\mathrm{UB}}(\Delta t)$, which quantify the lower and upper bounds of the maximum shiftable charging energy within a window of duration $\Delta t$. These indicators characterize the potential for load shifting through charging-side actions.

    \item \emph{Bidirectional extractable energy bounds} $E_{j,\mathrm{bi}}^{\mathrm{LB}}$ and $E_{j,\mathrm{bi}}^{\mathrm{UB}}$, which define the lower and upper bounds of the maximum extractable discharge energy. Only the discharged energy is considered flexible, since the subsequent recharge is required to restore the EV to the target $\mathrm{SoC}_{\mathrm{fin}}$ and therefore does not constitute a net grid-service contribution. The achievable bounds depend jointly on the available idle time and on battery-protection constraints.
\end{itemize}

\paragraph{Depth and temporal structure indicators}
\begin{itemize}
    \item \emph{Minimum reachable state of charge} $\mathrm{SoC}_{j,\min}^*$, which captures the maximum admissible depth of discharge under time-feasibility and battery-protection constraints. This indicator reveals whether flexibility is limited by idle-time availability or by conservative battery safeguards.
    \item \emph{Discharge and recharge durations} $T_{\mathrm{dis},j}$ and $T_{\mathrm{ch},j}$, which describe how the available idle time is allocated within a bidirectional maneuver and highlight asymmetries between discharge and recovery phases imposed by the BMS.
\end{itemize}

\paragraph{Feasibility indicator}
\begin{itemize}
    \item \emph{Feasibility rate} $\rho$, defined as the fraction of baseline operating points $\mathrm{SoC}_{j,k}$ for which a feasible bidirectional maneuver exists for a given $T_{\mathrm{idle}}$. This metric provides a probabilistic view of flexibility availability across the operating range of the EV.
\end{itemize}

Together, these indicators provide a compact yet comprehensive description of EV flexibility, explicitly linking time availability, energy magnitude, discharge depth, and feasibility. This set of metrics forms the basis for the sensitivity analyses, bound characterizations, and fleet-level assessments presented in the following section.



\section{Results: Charging Performance and Flexibility Potential} \label{sec:results}

This section presents the numerical results obtained by applying the proposed BMS-aware modeling and flexibility framework to the full set of EV charging sessions. The analysis is structured to progressively move from baseline charging performance to flexibility potential. First, the interaction between EV-specific Power--SoC characteristics and different DC charger ratings is evaluated to quantify effective charging time and delivered power. The section then examines unidirectional and bidirectional flexibility indicators, highlighting how time availability, energy bounds, and battery-protection constraints jointly shape the achievable flexibility at both single-vehicle and
fleet levels.


\subsection{BMS-aware charging performance across charger power levels}

Building on the Power--SoC and time-domain profiles, this section assesses how different fast and ultra-fast charger ratings interact with the BMS-aware charging limits of each EV. The goal is to quantify the effective charging performance that can be obtained when an EV with a given Power--SoC characteristic is connected to chargers of different nominal power levels.

A set of nominal DC charger ratings is considered:
\begin{equation}
    P_m^{\mathrm{CS}} \in \{7, 11, 22, 50, 60, 90, 120, 150, 180, 480\}~\text{kW},
\end{equation}
indexed by \(m = 1,\dots,10\). Each value \(P_m^{\mathrm{CS}}\) represents the maximum power that a given charging station can supply to an EV.

For each vehicle \(j\), two quantities are taken from the dataset introduced in Section~\ref{subsec:data}:
\begin{itemize}
    \item the battery capacity \(E_j\) (in kWh); and
    \item a discrete, BMS-aware Power--SoC curve \(P_j^{\mathrm{EV}}(s)\), defined for integer SoC values \(s \in \{1,\dots,100\}\), obtained from the interpolated profiles in \figurename~\ref{fig:EVPowVsSoC}.
\end{itemize}

The charging process is simulated in discrete time with sampling interval~\(t_{\mathrm{s}}\). In general, \(t_{\mathrm{s}}\) could be chosen according to the temporal resolution of the available data; in this study, it is fixed to \(t_{\mathrm{s}} = 1~\text{min}\). For each pair \((m,j)\)---charger rating \(P_m^{\mathrm{CS}}\) and EV \(j\)---the SoC trajectory starts from
\[
\mathrm{SoC}_j(0) = 1\%.
\]

At each time step \(k\), the instantaneous charging power is obtained by limiting the EV’s BMS-requested power with the charger rating:
\begin{equation}
    P_j(k) = \min\bigl(P_j^{\mathrm{EV}}(\mathrm{SoC}_j(k)),\; P_m^{\mathrm{CS}}\bigr).
\end{equation}

The SoC dynamics follow directly from the energy balance over one sampling interval:
\begin{equation}
   \mathrm{SoC}_j(k+1) = \mathrm{SoC}_j(k) + t_{\mathrm{s}}\,\frac{P_j(k)}{E_j}\,100, 
\end{equation}
and the simulation proceeds until \(\mathrm{SoC}_j(k) = 100\%\). The resulting sequences:
\[
\{P_j(k)\}_{k=0}^{K_{m,j}-1}
\quad\text{and}\quad
\{\mathrm{SoC}_j(k)\}_{k=0}^{K_{m,j}}
\]
are recorded, where \(K_{m,j}\) denotes the number of time steps required for EV \(j\) to reach 100\% SoC under charger rating \(P_m^{\mathrm{CS}}\).

For each pair \((m,j)\) we extract two performance indicators:
\begin{itemize}
    \item the total charging time from 1\% to 100\% SoC,
    \begin{equation}
        T_{m,j} = K_{m,j}\,t_{\mathrm{s}},
    \end{equation}
    expressed in minutes; and
    \item the average charging power over the full session,
    \begin{equation}
        \overline{P}_{m,j} = \frac{1}{K_{m,j}} \sum_{k=0}^{K_{m,j}-1} P_j(k).
    \end{equation}
\end{itemize}

By repeating this procedure for all charger ratings and all EVs, we obtain a comprehensive, BMS-aware comparison of charging time and mean delivered power across multiple charger technologies and EV categories.


\subsubsection{Charging Time}

In practical operation, EVs are rarely charged in the full 1\%--100\% SoC range. Most charging sessions occur within a mid-range window, typically between 20\% and 80\% SoC. To better reflect this behavior and to reduce the impact of interpolation artifacts near the extremes of the reconstructed profiles, a second analysis is performed for partial charging events from 20\% to 80\% SoC. This enables a more representative comparison of charging performance across charger power levels.

\begin{figure}
 \centering
 \includegraphics[width=\columnwidth]{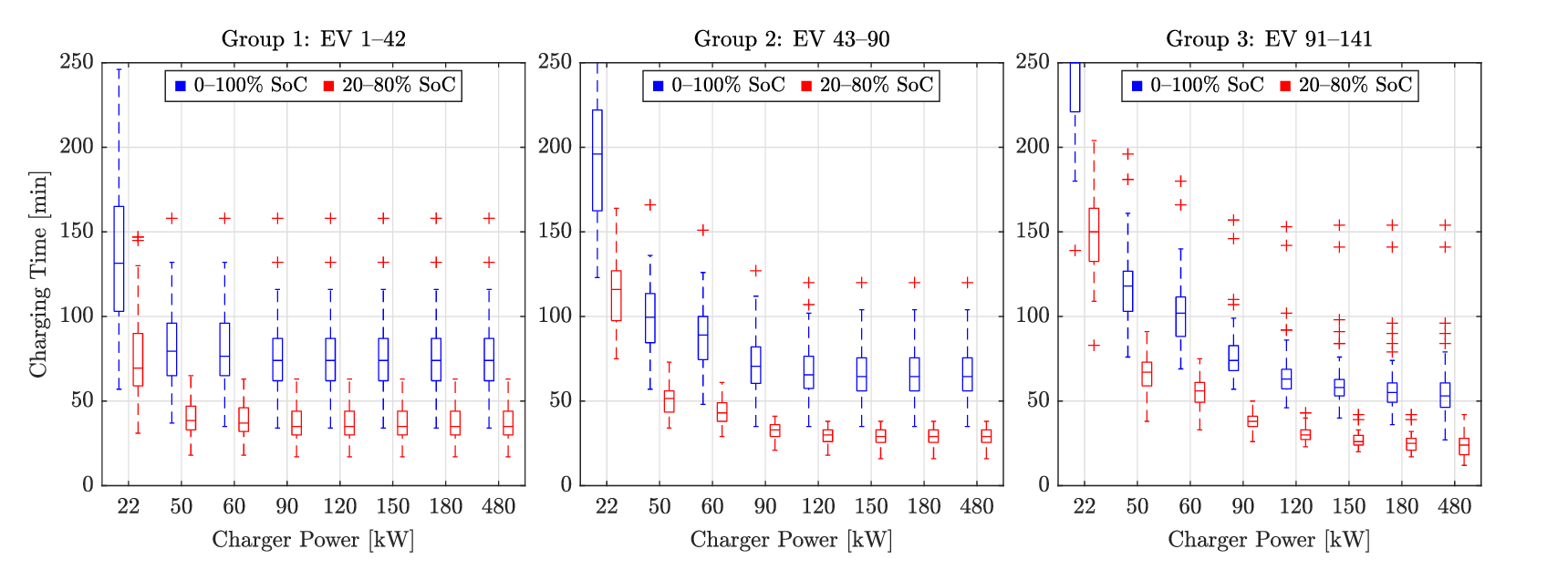}
 \caption{Charging-time comparison across charger ratings and EV groups. For each charger power level, boxplots report the distribution of charging time to reach 100\% SoC from 1\% (blue) and the time to move from 20\% to 80\% SoC (red).}
 \label{fig:TimeChAll}
 \end{figure}

\figurename~\ref{fig:TimeChAll} summarizes the charging-time results for the three EV groups presented in subsection~\ref{subsec:EVprof}. For each charger rating, the figure reports boxplots of the time required to reach 100\% SoC starting from 1\% (blue) and the time to move from 20\% to 80\% SoC (red). Across all groups, the 20\%--80\% window results in significantly shorter and less dispersed charging times. This reduction is particularly pronounced at high-power chargers, where charging beyond 80\% SoC is strongly limited by the BMS, leading to a marked increase in charging duration for the final portion of the session.

Low-power chargers (7~kW and 11~kW) are omitted from the boxplots to avoid compressing the vertical scale. At these power levels, charging durations are one order of magnitude larger than those observed for fast and ultra-fast chargers. As a reference, Group~1 EVs require approximately 2.5--12.5~h to charge from 1\% to 100\% at 7~kW, and 1.7--8~h at 11~kW; the corresponding 20\%--80\% times range between 1.5--7.5~h and 1--5~h, respectively. Groups~2 and~3 exhibit even longer durations due to their larger battery capacities. These results highlight the limited suitability of low-power chargers for applications requiring high energy throughput and fast turnaround.

\begin{figure}
 \centering
 \includegraphics[width=\columnwidth]{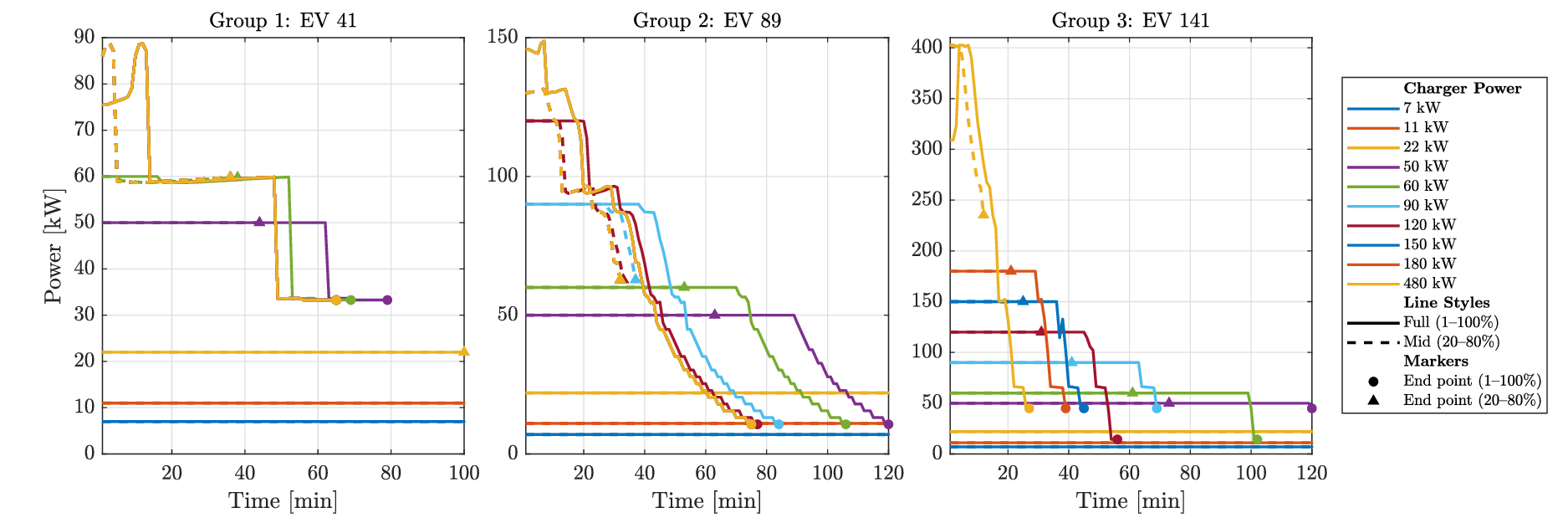}
 \caption{Example charging power profiles for one representative EV from each group, across all considered charger power levels. The curves show both the full 1\%--100\% charging trajectory and the 20\%--80\% operating window.}
 \label{fig:exampleEV}
 \end{figure}

The behavior underlying these distributions is illustrated in \figurename~\ref{fig:exampleEV}, which shows representative charging sessions for one EV from each group under all charger power levels. For each case, both the full 1\%--100\% trajectory and the 20\%--80\% interval are highlighted. Group~1 EVs exhibit moderate and sustained power levels, resulting in longer overall charging durations. Group~2 EVs achieve higher initial power, leading to faster mid-range SoC transitions. Group~3 ultra-fast-charging EVs display aggressive initial power intake followed by strong tapering at high SoC due to BMS-imposed limits.

For clarity, the trajectories corresponding to 7~kW, 11~kW, and, 22~kW chargers are truncated in the figure, as their extended duration would compress the time axis and obscure the dynamics of fast and ultra-fast charging.

\subsubsection{Mean Power}

While charging time quantifies how quickly an EV reaches a target SoC, the average delivered power reflects how effectively a given charger rating is utilized under BMS constraints. To capture this interplay, the mean charging power \(\overline{P}_{m,j}\) is computed for all charger–EV combinations, both over the full 1\%--100\% cycle and over the 20\%--80\% interval.

The resulting mean-power indicators are organized into two charger--EV matrices of size $10 \times 141$. The first matrix contains, for every charger rating and every EV, the average charging power computed over the full 1\%--100\% cycle. The second matrix reports the corresponding average power computed within the 20\%--80\% SoC interval, providing a more representative measure of charger utilization under typical fast-charging conditions.

\begin{figure}[t]
     \centering
     \includegraphics[width=\linewidth]{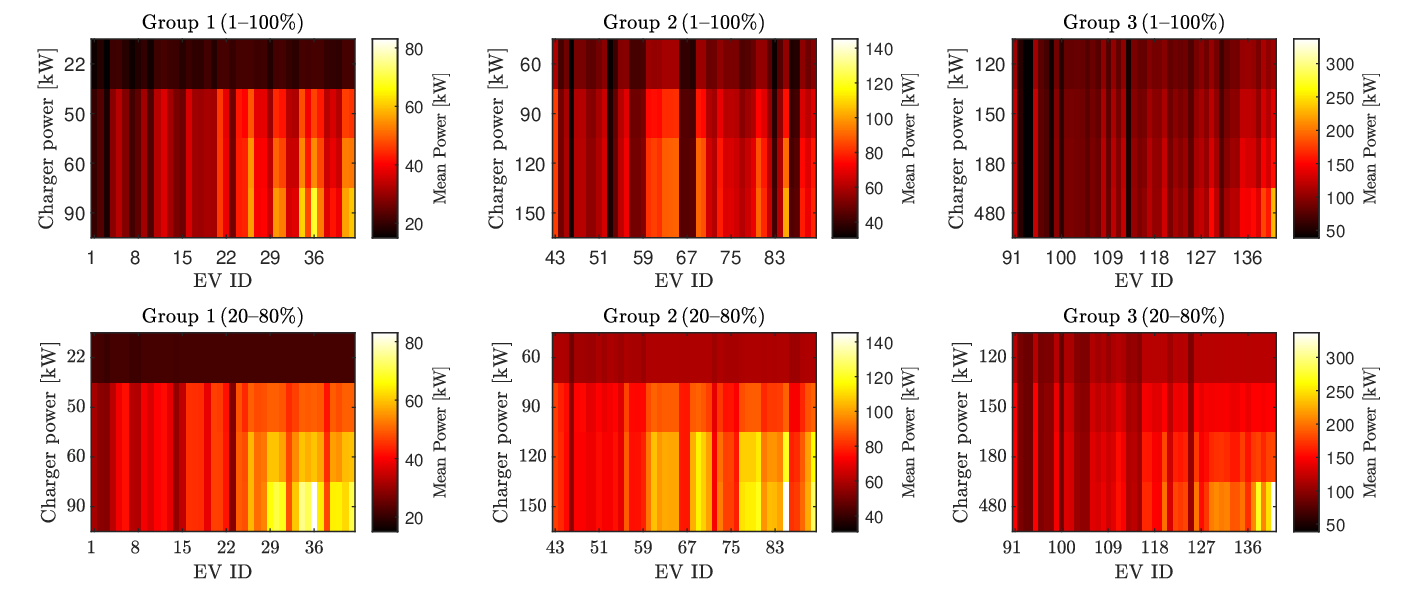}
     \caption{Mean charging power across charger ratings and EV groups. The top panel shows the average power over the full 1\%--100\% charging cycle, while the bottom panel reports the mean power over the 20\%--80\% SoC interval.}
     \label{fig:MeanPower}
 \end{figure}

\figurename~\ref{fig:MeanPower} presents these results as a set of heatmaps. The columns correspond to EV IDs (grouped as in the previous sections) and the rows to charger ratings. The top panel shows the mean power over the full 1\%--100\% charge, while the bottom panel reports the mean power over the 20\%--80\% SoC interval. For each EV group, the color scale is shared between panels, allowing a direct comparison between full and partial charging conditions.

Several key trends emerge. First, at low charger ratings, the mean power closely matches the nominal value, indicating that the charger is the limiting factor and the BMS does not constrain the charging process. As the charger rating increases, the mean power progressively saturates at values well below the station rating for many EVs, particularly in Group~1. This behavior reveals diminishing returns from increasing charger power, as the BMS becomes the dominant limiting factor. Second, restricting the analysis to the 20\%--80\% SoC interval consistently increases the mean power, since the low-power tail near full charge is excluded. This confirms that mid-range SoC operation is the most effective region for exploiting fast and ultra-fast charging capabilities. These results highlight that the effective utilization of high-power charging infrastructure is strongly dependent on the EV-specific Power--SoC characteristics, which directly shape both charging performance and the temporal structure of flexibility.


\subsection{Characterization of Discharge Capability}
\label{subsec:discharge_analysis}

Before quantifying flexibility, it is necessary to characterize the intrinsic discharge capability of the EV fleet in the time domain. This subsection analyzes the synthetic discharge power--SoC profiles introduced in Section~\ref{subsec:synthetic_discharge}, providing baseline indicators on discharge duration and mean power levels.

For each EV, a full discharge event is reconstructed starting from a fully charged state $\mathrm{SoC}_j(0)=100\% $ and evolving in discrete time with sampling interval $t_{\mathrm{s}}$ according to the energy balance
\begin{equation}
\mathrm{SoC}_j(k+1)=
\mathrm{SoC}_j(k)-t_{\mathrm{s}}
\frac{P^{\mathrm{dis}}_j\!\left(\mathrm{SoC}_j(k)\right)}{C_j}\,100,
\label{eq:soc_discharge}
\end{equation}
where $C_j$ denotes the battery capacity and $P^{\mathrm{dis}}_j(\cdot)$ is the discharge power--SoC characteristic defined on the discrete SoC. The trajectory terminates when $\mathrm{SoC}_j(k)=1\%$.

The resulting time-domain sequences $\{\mathrm{SoC}_j(k)\}$ and $\{P^{\mathrm{dis}}_j(k)\}$ reveal pronounced heterogeneity across vehicles, reflecting differences in battery size and BMS discharge limits. This variability directly affects both the depth and the duration of admissible bidirectional maneuvers.

\subsubsection{Discharge duration}

For each EV, the total discharge time is defined as
\begin{equation}
T^{\mathrm{dis}}_j = K^{\mathrm{dis}}_j\, t_{\mathrm{s}},
\end{equation}
where $K^{\mathrm{dis}}_j$ denotes the number of time steps required to reach the lower SoC bound.

\figurename~\ref{fig:DischargeTimeMeanP}(a) reports the distribution of discharge durations for the full $100\%\!\rightarrow\!1\%$ event (blue) and for the mid-range $80\%\!\rightarrow\!20\%$ window (red), grouped by EV category. As expected, the $80\%\!\rightarrow\!20\%$ durations are systematically shorter. Across all groups, the spread of the distributions highlights a significant heterogeneity in discharge persistence, directly linked to the variability of the discharge power--SoC characteristics.

\subsubsection{Mean discharge power}

To complement the time-based analysis, the average discharge power is computed as
\begin{equation}
\overline{P}^{\mathrm{dis}}_j =
\frac{1}{K^{\mathrm{dis}}_j}
\sum_{k=0}^{K^{\mathrm{dis}}_j-1}
P^{\mathrm{dis}}_j(k).
\end{equation}

\begin{figure}
 \centering
 \includegraphics[width=\columnwidth]{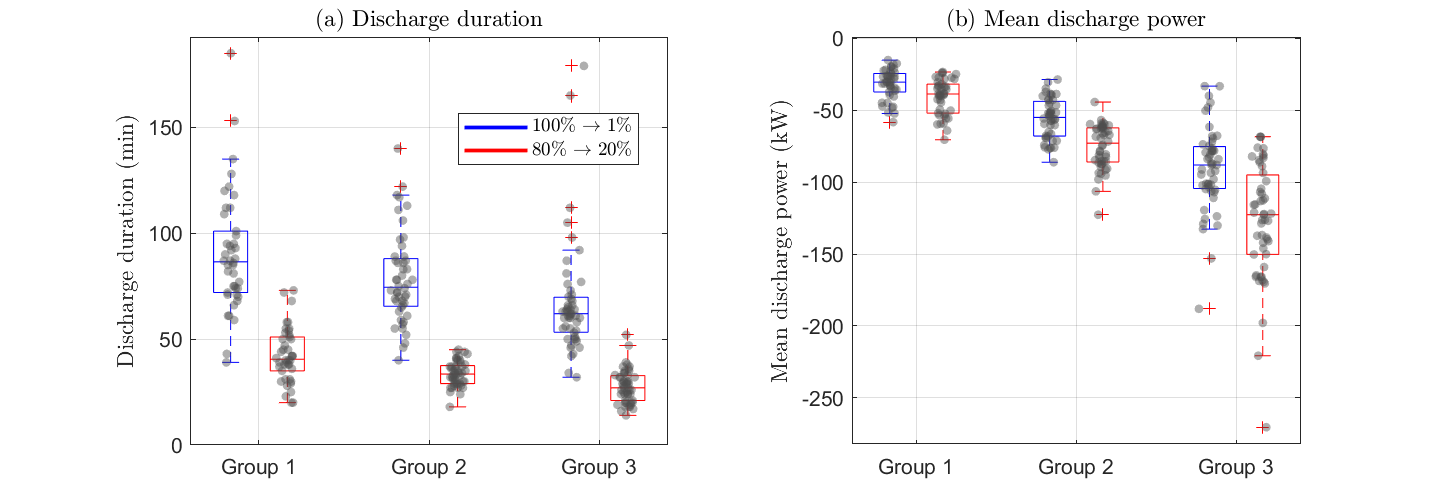}
 \caption{Time-domain characterization of EV discharge capability across groups. (a) Discharge duration for the full $100\%\!\rightarrow\!1\%$ event (blue) and the $80\%\!\rightarrow\!20\%$ window (red). (b) Corresponding mean discharge power over the same intervals.}
 \label{fig:DischargeTimeMeanP}
 \end{figure}

\figurename~\ref{fig:DischargeTimeMeanP}(b) shows that the mean discharge power increases from Group~1 to Group~3, reflecting the higher power capability of larger and ultra-fast-charging EVs. Additionally, focusing on the $80\%\!\rightarrow\!20\%$ interval generally results in higher average power values.

\subsection{Unidirectional Energy-Based Flexibility Assessment}
\label{subsec:flex_analysis}

\begin{figure}
     \centering
     \includegraphics[width=\linewidth]{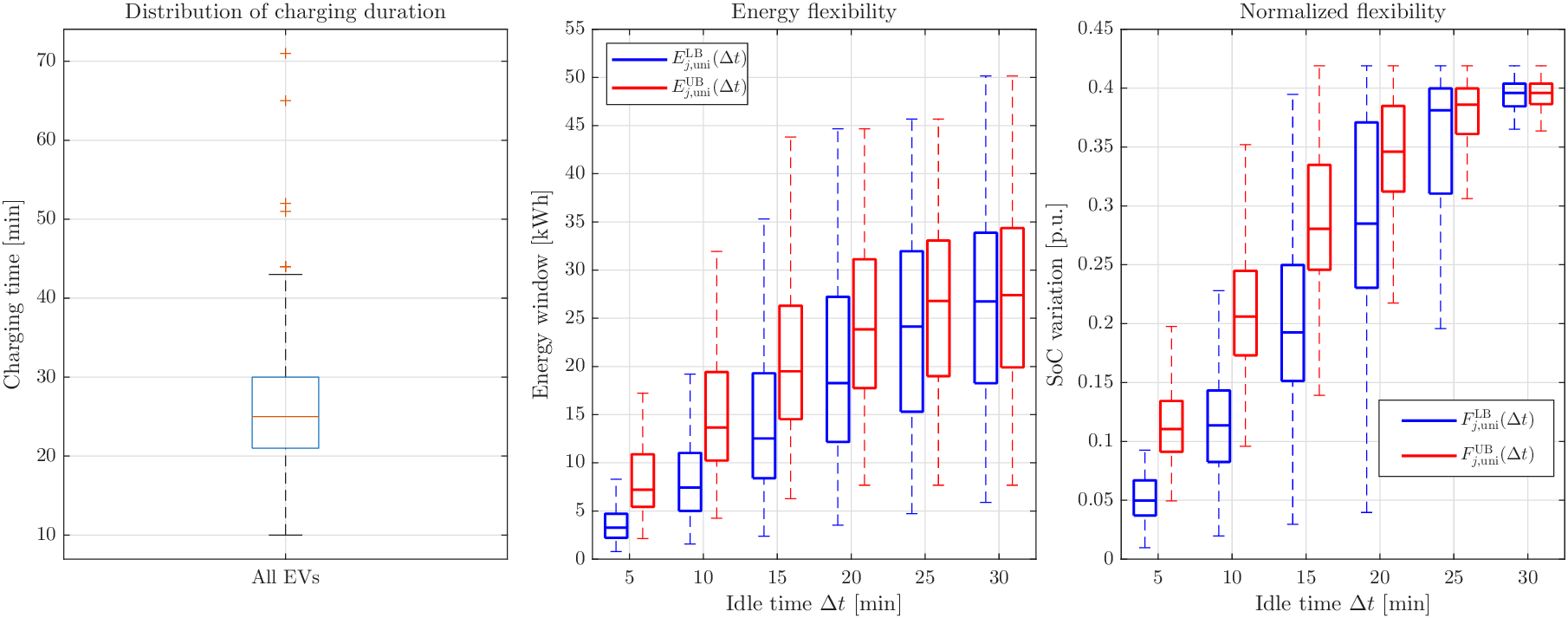}
     \caption{Unidirectional flexibility assessment across the fleet. (a) Distribution of charging durations $T_j$ required to progress from approximately 50\% to 90\% SoC. (b) Boxplots of the lower and upper bounds of the maximum shiftable energy, $E_{j,\mathrm{uni}}^{\mathrm{LB}}(\Delta t)$ and $E_{j,\mathrm{uni}}^{\mathrm{UB}}(\Delta t)$, for different window lengths $\Delta t$. (c) Corresponding normalized flexibility in SoC terms, $F_{j,\mathrm{uni}}^{\mathrm{LB}}(\Delta t)$ and $F_{j,\mathrm{uni}}^{\mathrm{UB}}(\Delta t)$. Larger windows lead to wider feasible flexibility ranges.}
     \label{fig:FlexibilityRes}
 \end{figure}

The unidirectional flexibility framework is applied to all DC charging sessions to quantify how much of the BMS-constrained charging power can be temporally shifted without affecting the final SoC. To ensure a consistent comparison across heterogeneous charging behaviors, the analysis focuses on an SoC interval approximately spanning from $50\%$ to $90\%$. For each vehicle, the corresponding duration $T_j$, i.e., the minimum time required to complete this transition, is computed. The distribution of $T_j$, reported in \figurename~\ref{fig:FlexibilityRes}\,(a), ranges from approximately 10 to 70 minutes, with a median around 25 minutes. This spread reflects the diversity of BMS behavior at high SoC. During this interval, all delivered energy is required to reach the target SoC, and therefore no flexibility is available. Flexibility only emerges when the connection time exceeds $T_j$.

For each vehicle and for window lengths $\Delta t\in\{5,10,15,20,25,30\}$\,min, the sliding-window analysis yields the bounds
\[
E_{j,\mathrm{uni}}^{\mathrm{LB}}(\Delta t), \qquad
E_{j,\mathrm{uni}}^{\mathrm{UB}}(\Delta t),
\]
which define the lower and upper bounds of the maximum shiftable energy within a $\Delta t$-minute interval. \figurename~\ref{fig:FlexibilityRes}\,(b) shows the aggregated distributions across the fleet. As expected, both bounds increase with $\Delta t$, while their separation widens, indicating a growing potential for intra-session power redistribution as the available time window increases.

To remove the influence of battery capacity, the normalized metrics
\[
F_{j,\mathrm{uni}}^{\mathrm{LB}}(\Delta t), \qquad
F_{j,\mathrm{uni}}^{\mathrm{UB}}(\Delta t)
\]
are reported in \figurename~\ref{fig:FlexibilityRes}\,(c). Even short windows (5--10\,min) allow for several percentage points of SoC variation across a significant portion of the fleet. For $\Delta t \ge 20$\,min, the achievable SoC variation increases markedly, showing that fast and ultra-fast charging sessions can provide meaningful short-term flexibility when sufficient idle time is available.

These results define the unidirectional flexibility bounds, which serve as the baseline for the bidirectional flexibility analysis presented next.


\subsection{Bidirectional flexibility under idle-time constraints}
\label{subsec:bidirectional_results}

This subsection presents the results of the bidirectional flexibility framework, focusing on how idle-time availability shapes the amount of energy that can be extracted and later recovered within a charging session. The analysis is structured in two steps. First, the sensitivity of bidirectional flexibility to idle time is examined for representative EVs, providing an intuitive interpretation of feasibility, temporal realization, and extractable energy. Second, the framework is extended to a heterogeneous fleet to characterize aggregate trends and variability.

\subsubsection{Impact of idle time on single-EV bidirectional flexibility}
\label{subsec:idle_sensitivity}

To illustrate the underlying mechanisms, two representative EVs are selected: one fast-charging EV and one ultra-fast-charging EV. The analysis is organized into three complementary perspectives: (i) \emph{feasibility}, identifying where bidirectional operation is possible in the $(\mathrm{SoC}_{j,k}, T_{\mathrm{idle}})$ domain; (ii) \emph{temporal realization}, describing how the maneuver unfolds over time; and (iii) \emph{energy bounds}, quantifying the extractable discharge energy.

\begin{figure}
 \centering
 \includegraphics[width=\linewidth]{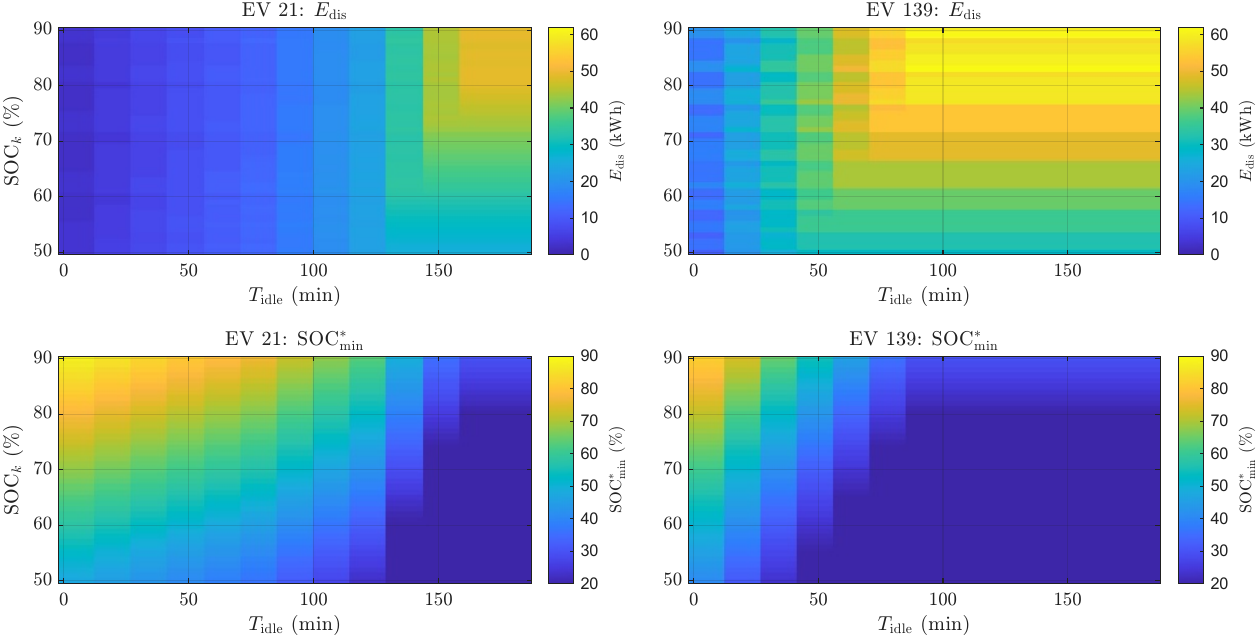}
 \caption{Bidirectional flexibility maps for two representative EVs (fast-charging and ultra-fast-charging). Top row: extractable discharge energy $E_{j,\mathrm{dis}}(\mathrm{SoC}_{j,k}, T_{\mathrm{idle}})$. Bottom row: corresponding minimum reachable state of charge $\mathrm{SoC}_{j,\min}^{*}(\mathrm{SoC}_{j,k}, T_{\mathrm{idle}})$.}
 \label{fig:BiFlex_PanelA}
 \end{figure}

\figurename~\ref{fig:BiFlex_PanelA} addresses feasibility by showing the discharge-energy maps $E_{j,\mathrm{dis}}(\mathrm{SoC}_{j,k}, T_{\mathrm{idle}})$. These maps highlight the regions where bidirectional operation becomes energetically relevant. As idle time increases, the feasible region expands, allowing either longer discharge durations or deeper discharge levels while still ensuring recovery to $\mathrm{SoC}_{\mathrm{fin}}$. The corresponding maps of $\mathrm{SoC}_{j,\min}^{*}(\mathrm{SoC}_{j,k}, T_{\mathrm{idle}})$ reveal the dominant limiting mechanism: for short idle times, feasibility is constrained by time availability, whereas for larger idle times, the minimum SoC constraint becomes binding. Lower admissible $\mathrm{SoC}_{j,\min}^{*}$ values consistently correspond to higher extractable energy.

\begin{figure}
 \centering
 \includegraphics[width=\linewidth]{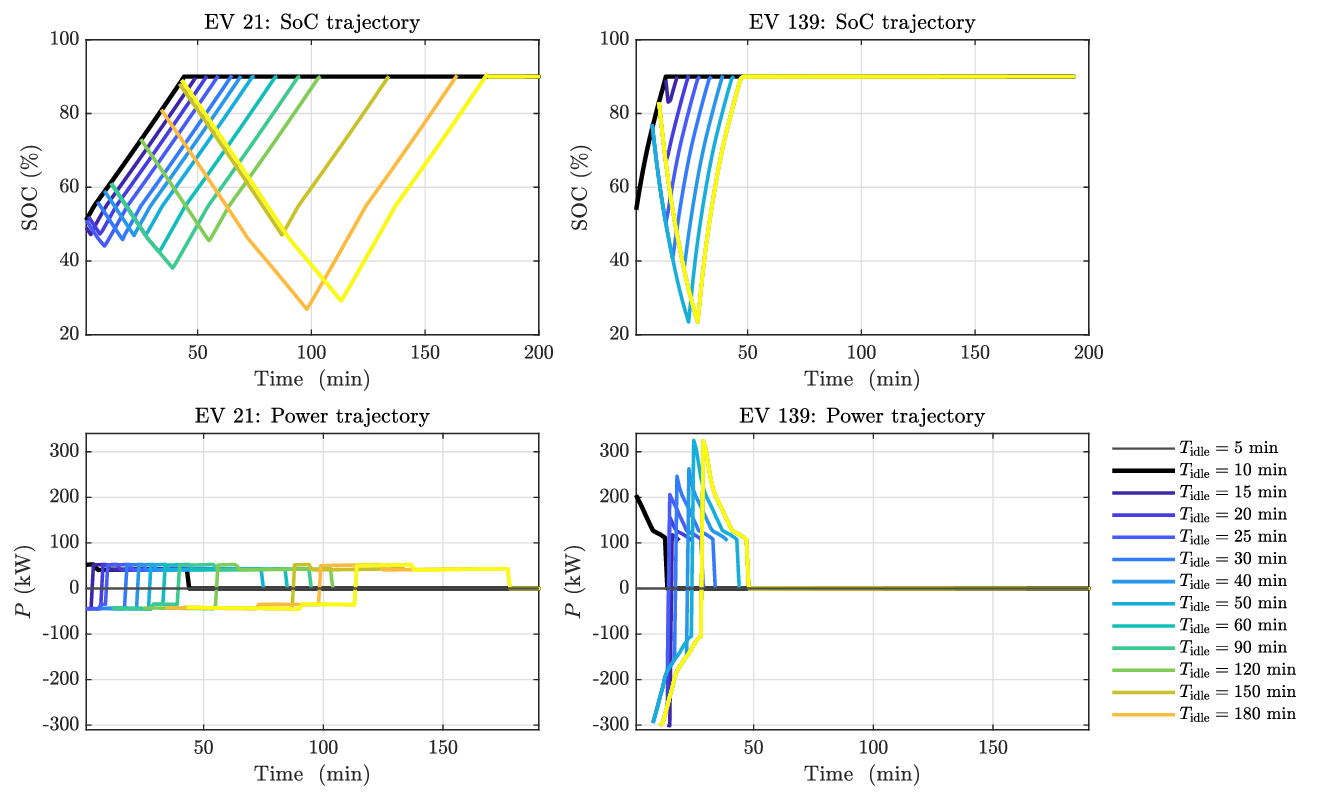}
 \caption{Time-domain trajectories associated with the upper-bound bidirectional maneuver for two representative EVs.}
 \label{fig:BiFlex_PanelB}
 \end{figure}

The temporal realization of these results is illustrated in \figurename~\ref{fig:BiFlex_PanelB}, which shows the connection trajectories associated with the upper-bound solution for each value of $T_{\mathrm{idle}}$. The $\mathrm{SoC}(t)$ profiles exhibit a characteristic dip-and-recovery behavior around the baseline charging trajectory, while the corresponding $P(t)$ profiles separate discharge and recharge phases. Differences across EVs reflect their intrinsic charging and recovery dynamics, with ultra-fast-charging vehicles completing the same energy exchange over shorter time horizons.

\begin{figure}
 \centering
 \includegraphics[width=0.8\linewidth]{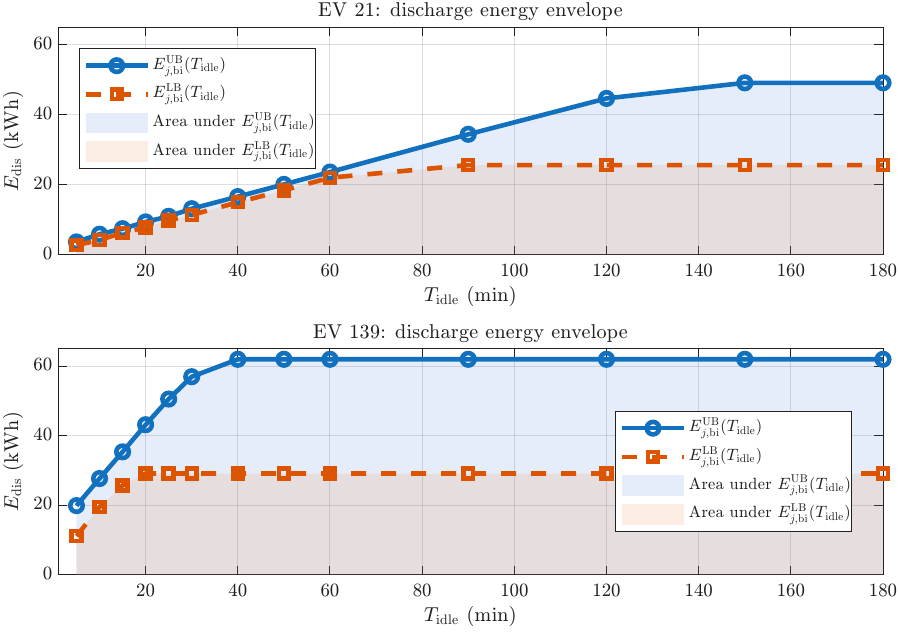}
 \caption{Discharge-energy envelopes for two representative EVs as a function of idle time $T_{\mathrm{idle}}$.}
 \label{fig:BiFlex_PanelC}
 \end{figure}

Finally, \figurename~\ref{fig:BiFlex_PanelC} summarizes the results through discharge-energy bounds as a function of idle time. For each EV, the upper curve corresponds to $E_{j,\mathrm{bi}}^{\mathrm{UB}}(T_{\mathrm{idle}})$, i.e., the maximum extractable discharge energy across all feasible initiation points $\mathrm{SoC}_{j,k}$, while the lower curve represents $E_{j,\mathrm{bi}}^{\mathrm{LB}}(T_{\mathrm{idle}})$, i.e., the lower bound of the maximum achievable energy. These envelopes provide a compact representation of bidirectional flexibility and highlight its strong dependence on idle time and operating conditions.

\subsubsection{Fleet-level bidirectional flexibility across idle times}
\label{subsec:fleet_multiEV_multiIdle}

\begin{figure}
 \centering
 \includegraphics[width=0.5\linewidth]{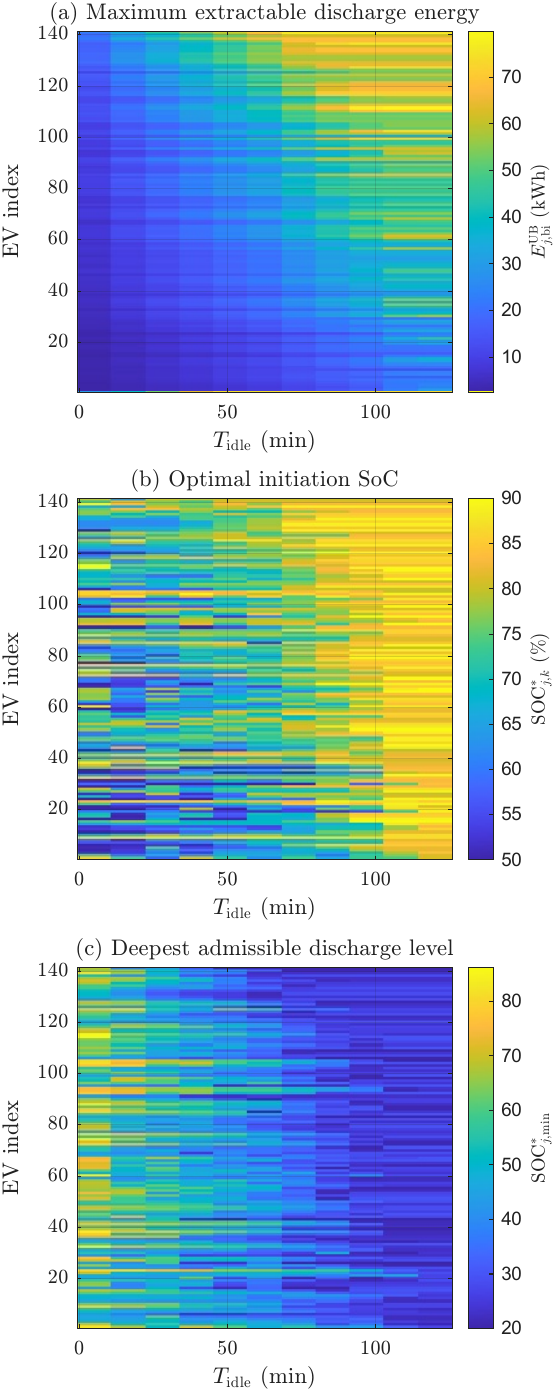}
 \caption{Fleet-level characterization of bidirectional flexibility as a function of idle time. Left: upper bound of the maximum extractable discharge energy $E_{j,\mathrm{bi}}^{\mathrm{UB}}$. Center: optimal initiation point $\mathrm{SoC}_{j,k}^{*}$ at which $E_{j,\mathrm{bi}}^{\mathrm{UB}}$ is attained. Right: corresponding minimum reachable state of charge $\mathrm{SoC}_{j,\min}^{*}$.}
 \label{fig:Fleet_Heatmaps}
 \end{figure}

The bidirectional flexibility framework is extended to a heterogeneous EV fleet to assess aggregate trends and inter-vehicle variability as a function of idle time. \figurename~\ref{fig:Fleet_Heatmaps} summarizes the fleet-level behavior through three complementary heatmaps.

The first heatmap reports $E_{j,\mathrm{bi}}^{\mathrm{UB}}(T_{\mathrm{idle}})$, i.e., the upper bound of the maximum extractable discharge energy for each EV. The results reveal pronounced heterogeneity across the fleet: some vehicles exhibit limited bidirectional potential even for large idle times, while others can extract substantially higher energy as $T_{\mathrm{idle}}$ increases. The observed saturation for large $T_{\mathrm{idle}}$ indicates a transition from time-limited to battery-limited operation.

The second heatmap reports the optimal initiation point $\mathrm{SoC}_{j,k}^{*}(T_{\mathrm{idle}})$, i.e., the state of charge at which the upper-bound energy is achieved. As idle time increases, the optimal point shifts toward higher SoC levels, enabling longer or deeper discharge trajectories while still ensuring recovery to $\mathrm{SoC}_{\mathrm{fin}}$ within the available connection time.

The third heatmap shows the corresponding minimum reachable state of charge $\mathrm{SoC}_{j,\min}^{*}(T_{\mathrm{idle}})$ at the upper-bound operating point. For short idle times, $\mathrm{SoC}_{j,\min}^{*}$ is constrained by time feasibility, whereas for larger idle times it saturates at battery-protection limits, including the imposed 60\% depth-of-discharge constraint.

\begin{figure}
 \centering
 \includegraphics[width=0.85\linewidth]{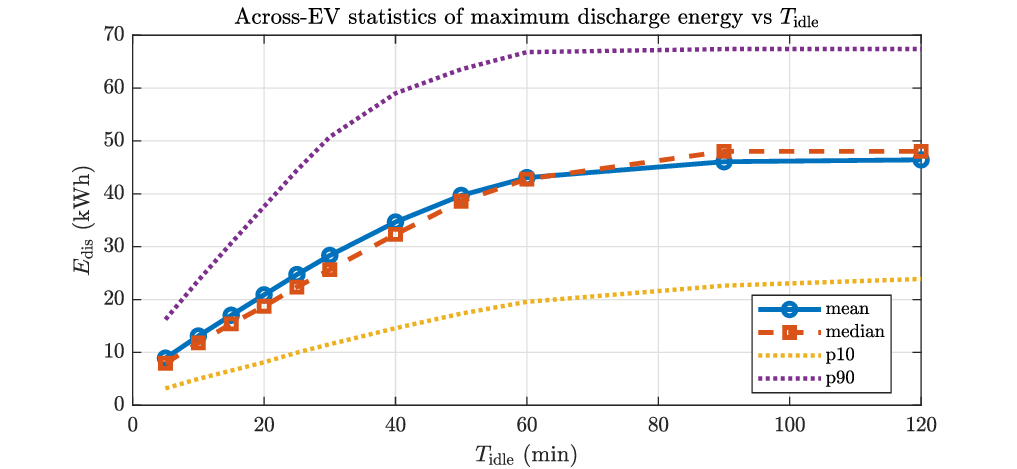}
 \caption{Across-EV statistics of the upper bound of the maximum extractable discharge energy $E_{j,\mathrm{bi}}^{\mathrm{UB}}$ as a function of idle time $T_{\mathrm{idle}}$.}
 \label{fig:Fleet_Statistics}
 \end{figure}

A compact statistical synthesis is provided in \figurename~\ref{fig:Fleet_Statistics}, which reports the mean, median, and inter-quantile (10--90\%) range of $E_{j,\mathrm{bi}}^{\mathrm{UB}}$ across EVs as a function of idle time. The monotonic increase of these metrics confirms that idle-time availability is the dominant enabler of bidirectional flexibility, while the widening inter-quantile band highlights increasing dispersion across heterogeneous EVs.

\section{Discussion: Implications for Grid Operation and Energy Services} \label{sec:discussion}

The results show that flexibility from fast and ultra-fast EV charging cannot be treated as power-controllable. Instead, it is fundamentally \emph{energy-constrained} and \emph{state-dependent}. Although the charging station may be capable of delivering high power levels, the actual charging trajectory is governed by the EV BMS and its associated battery constraints. Consequently, instantaneous charging power cannot be arbitrarily modulated, and flexibility must be interpreted as a bounded energy-shifting capability defined over the SoC trajectory and the available connection time.

From an operational perspective, this implies that EVs connected to fast chargers cannot be modeled as continuously dispatchable resources. Rather, they should be treated as \emph{time-constrained energy buffers}, whose flexibility is activated within feasible energy bounds. These bounds depend on the SoC, the BMS-imposed Power--SoC profile, and the remaining connection time.
These limits correspond to the lower and upper bounds of the maximum achievable flexibility, which define the feasible operating region for any admissible charging or discharging trajectory.

A key result is that flexibility is primarily driven by \emph{idle time}, rather than charger rating. While higher charger power reduces charging duration, it simultaneously reduces the available window for flexibility. As a result, flexibility availability is more strongly governed by user connection behavior than by infrastructure capacity.
This suggests that strategies aiming to extract value from EV flexibility should prioritize \emph{connection management} (e.g., delayed unplugging, parking incentives) over purely increasing charger power.

Bidirectional flexibility further expands the feasible energy region, but remains \emph{bounded and feasibility-limited}. Rather than corresponding to a single operating pattern, it defines a set of admissible trajectories within the bounds of the maximum achievable energy exchange. These limits are jointly determined by time availability and battery-protection constraints, implying that energy injection to the grid is only possible within specific operating regions and cannot be sustained continuously.

Battery-protection constraints emerge as a dominant limiting factor. Minimum SoC thresholds and depth-of-discharge limits significantly reduce the exploitable flexibility, even when time availability would allow deeper energy exchange. This highlights that any practical deployment of EV-based energy services must explicitly account for degradation-aware constraints.
This indicates that flexibility is not only time-limited, but also structurally constrained by battery usage policies.

These findings have direct implications for energy service design. First, fast-charging EVs are inherently suited for \emph{short-duration, energy-limited services}, such as peak shaving, congestion management, or fast reserve activation over limited time windows. Second, flexibility should be scheduled using \emph{energy-aware optimization}, rather than power setpoint control, ensuring that all trajectories remain within feasible SoC and time constraints. Third, aggregation frameworks must explicitly incorporate \emph{state-dependent availability}, where flexibility can only be activated under favorable SoC and idle-time conditions, rather than assuming continuous controllability.

In addition, the results indicate that the mid-range SoC window (e.g., 20--80\%) provides the most effective region for flexibility exploitation, as it combines higher power levels with less restrictive BMS limitations. Energy services targeting fast-charging EVs should therefore be explicitly designed around this operating region.

Finally, the proposed framework enables practical flexibility quantification without requiring direct V2G measurements. By combining measured charging data with physically consistent bounds, it provides realistic estimates of flexibility potential that can be directly integrated into planning and operational studies.

Overall, fast and ultra-fast EV charging should be modeled as a \emph{constrained and opportunistic source of flexibility}, whose value emerges only under specific combinations of SoC, connection time, and battery-protection limits.


\section{Conclusions}
\label{sec:conclusions}

This work presented a data-driven framework to quantify flexibility in fast and ultra-fast EV charging systems as an energy-bounded and time-constrained resource. The proposed formulation is grounded on real-world charging data, enabling the reconstruction and analysis of representative Power--SoC trajectories across a heterogeneous EV fleet. This data-driven foundation allows capturing the intrinsic variability and constraints of practical charging behavior.

Departing from conventional power-controllable assumptions, flexibility was formulated as the bounds of the maximum realizable energy that can be shifted or extracted under the joint constraints imposed by the Power--SoC characteristics, the available connection time, and battery-protection limits.

For unidirectional operation, flexibility was characterized through the upper and lower bounds of the maximum shiftable charging energy within finite time windows, revealing that load-shifting potential is fundamentally constrained by the temporal structure of the charging trajectory and the availability of idle time. For bidirectional operation, the framework identified the lower and upper bounds of the maximum extractable discharge energy, showing that feasible flexibility is inherently time-coupled and strongly dependent on the initiation point along the charging trajectory.

The results demonstrate that flexibility in fast charging systems is not a continuously dispatchable resource, but rather a constrained and state-dependent capability that can only be exploited within well-defined energy and time bounds. In particular, the lower bound represents a conservative and consistently achievable level of flexibility, while the upper bound defines the maximum potential that can be realized under favorable operating conditions.

At the fleet level, the proposed formulation enables the construction of flexibility bounds, feasibility maps, and energy indicators that explicitly account for EV heterogeneity and idle-time availability. These results provide a physically grounded basis for integrating EV flexibility into system-level analysis, supporting the design of energy services such as peak shaving, congestion management, and short-duration reserve provision. This enables a consistent transition from single-EV characterization to system-level flexibility assessment.

Overall, the proposed framework shifts the perspective from flexibility as a controllable action to flexibility as a realizable limit, providing a physically consistent and operationally meaningful representation of EV charging flexibility grounded in real-world data. Future work will focus on translating the proposed flexibility bounds into operational tools for real-time scheduling and energy service provision, as well as extending the framework to other electrified transport systems such as heavy-duty and port-based applications.

\appendix
\section{EV Brand–Model Groups}
\label{app1}

This appendix reports the EV brand–model groups used in the analysis. The grouping is based on publicly available brand/model compatibility information for DC fast chargers and reflects vehicles that exhibit similar charging capabilities and charging behavior. Only groups containing more than one model are included. Table~\ref{tab:ev_group} provides the complete list of groups and their associated vehicle models.

{\scriptsize
\begin{longtable}{c p{4cm} p{4cm} p{4cm}}
\caption{EV brand--model group used in the analysis. Only groups containing more than one model are shown.}%
\label{tab:ev_group}\\
\hline
Group & Brand -- Model & Brand -- Model & Brand -- Model \\
\hline
\endfirsthead

\hline
Group & Brand -- Model & Brand -- Model & Brand -- Model \\
\hline
\endhead

\hline
\endfoot

\hline
\endlastfoot

2   & Skoda -- Citigo e iV & Volkswagen -- e-Up! & Seat -- Mii Electric \\[3pt]

7   & Hyundai -- Kona Electric & Kia -- e-Niro & Kia -- e-Soul \\
7   & Hyundai -- Kona & & \\[3pt]

10  & LEVC -- VN5 & LEVC -- TX & \\[3pt]

33  & Hyundai -- Kona Electric & Kia -- e-Niro & Kia -- e-Soul \\
33  & Hyundai -- Kona & & \\[3pt]

39  & Fiat -- 500e & Abarth -- 500e & \\[3pt]

47  & Fiat -- e-Scudo & Fiat -- E-Ulysse & Opel -- Vivaro-e \\
47  & Opel -- Zafira-e & Peugeot -- e-Expert & Peugeot -- e-Traveller \\
47  & Toyota -- ProAce Electric & Toyota -- Proace Verso Electric & Citro\"en -- \"E-Jumpy \\[3pt]

48  & Toyota -- Proace Verso Electric & Vauxhall -- Vivaro-e & Toyota -- ProAce Electric \\[3pt]

49  & Audi -- Q4 e-tron Sportback 35 & Volkswagen -- ID.4 & Volkswagen -- ID.5 \\
49  & Audi -- Q4 e-tron 35 & & \\[3pt]

50  & Skoda -- Enyaq iV 60 & Skoda -- Enyaq Coup\'e iV 60 & \\[3pt]

51  & Citro\"en -- \"e-C4 & Citro\"en -- \"e-C4 X & Citro\"en -- \"E-Jumpy \\
51  & Citro\"en -- \"E-Spacetourer & DS -- 3 Crossback E-Tense & Fiat -- e-Dobl\`o \\
51  & Fiat -- e-Scudo & Fiat -- E-Ulysse & Toyota -- ProAce City Electric \\
51  & Toyota -- ProAce Electric & Toyota -- ProAce Verso Electric & Vauxhall -- Combo Electric \\
51  & Vauxhall -- Mokka-e & Vauxhall -- Vivaro-e & Opel -- Combo-e \\
51  & Opel -- Corsa-e & Opel -- Mokka-e & Opel -- Vivaro-e \\
51  & Opel -- Zafira-e & Peugeot -- e-2008 & Peugeot -- e-208 \\
51  & Peugeot -- e-Expert & Peugeot -- e-Partner & Peugeot -- e-Rifter \\
51  & Peugeot -- e-Traveller & Citro\"en -- \"E-Berlingo & \\[3pt]

54  & Renault -- 5 E-Tech & Nissan -- Micra & Renault -- 4 E-Tech \\[3pt]

57  & Opel -- Astra Sports Tourer Electric & Opel -- Corsa Electric & Opel -- Mokka Electric \\
57  & Opel -- Astra Electric & & \\[3pt]

58  & Peugeot -- e-308 SW & Peugeot -- e-308 & \\[3pt]

60  & Mercedes-Benz -- eVito & Mercedes-Benz -- eVito Tourer & Mercedes-Benz -- EQV 300 \\[3pt]

69  & Audi -- e-tron Sportback 50 Quattro & Audi -- e-tron 50 Quattro & \\[3pt]

72  & Volkswagen -- ID.3 Pro & Cupra -- Born 58 & \\[3pt]

76  & Renault -- Scenic E-Tech EV60 & Renault -- Megane E-Tech V60 & \\[3pt]

79  & Volvo -- EC40 & Volvo -- EX40 Recharge & Volvo -- XC40 Recharge Single Motor \\
79  & Volvo -- C40 Recharge & & \\[3pt]

82  & NIO -- EL7 & NIO -- ET5 & NIO -- ET5 Touring \\
82  & NIO -- ET7 & NIO -- EL6 & \\[3pt]

85  & NIO -- EL6 & NIO -- ET5 & NIO -- ET5 Touring \\
85  & NIO -- EL7 & & \\[3pt]

87  & Lexus -- RZ 300/450e & Toyota -- bZ4X & \\[3pt]

89  & Renault -- Scenic E-Tech EV87 & Nissan -- ARIYA & \\[3pt]

91  & Audi -- Q8 e-tron Sportback 50 quattro & Audi -- Q8 e-tron 50 quattro & \\[3pt]

92  & Volvo -- XC40 Recharge & Polestar -- 2 (78~kWh) & Volvo -- C40 Recharge \\[3pt]

93  & Ford -- Mach-E GT & Ford -- Mach-E ER & \\[3pt]

95  & Audi -- e-tron 55 Quattro & Audi -- e-tron Sportback 55 Quattro & \\[3pt]

97  & Smart -- \#3 & Smart -- \#1 & \\[3pt]

98  & Volvo -- EX30 Cross Country & Volvo -- EX30 & \\[3pt]

99  & Audi -- Q8 e-tron Sportback 55 quattro & Audi -- SQ8 e-tron & Audi -- SQ8 e-tron Sportback \\
99  & Audi -- Q8 e-tron 55 quattro & & \\[3pt]

103 & Volkswagen -- ID.4 & Volkswagen -- ID.4 GTX & Volkswagen -- ID.5 \\
103 & Volkswagen -- ID.5 GTX & Cupra -- Tavascan & Skoda -- Enyaq 85 \\
103 & Skoda -- Enyaq Coup\'e 85 & Volkswagen -- ID.3 Pro S & \\[3pt]

104 & Skoda -- Enyaq iV 80 / 85 & Volkswagen -- ID. Buzz & Volkswagen -- ID. Buzz Cargo \\
104 & Skoda -- Enyaq Coup\'e iV 80 & & \\[3pt]

105 & Audi -- Q4 e-tron 50/55 & Audi -- Q4 e-tron Sportback 50/55 & Skoda -- Enyaq Coup\'e iV 80 / 85 \\
105 & Skoda -- Enyaq iV 80 / 85 & Audi -- Q4 e-tron 40/45 & \\[3pt]

110 & BMW -- iX M60 & BMW -- iX 50 & \\[3pt]

116 & Mercedes-Benz -- EQS SUV & Mercedes-Benz -- EQS & \\[3pt]

121 & Tesla -- Model S Plaid & Tesla -- Model X & Tesla -- Model X Plaid \\
121 & Tesla -- Model S & & \\[3pt]

122 & Hyundai -- IONIQ 6 Long Range & Hyundai -- IONIQ 5 77 & \\[3pt]

123 & Porsche -- Taycan Cross Turismo & Porsche -- Taycan (79.2~kWh) & \\[3pt]

126 & Genesis -- GV70 & Kia -- EV6 77 & Genesis -- GV60 \\[3pt]

127 & Volvo -- EX90 & Polestar -- 3 & \\[3pt]

129 & Maserati -- GranTurismo Folgore & Maserati -- GranCabrio Folgore & \\[3pt]

130 & Kia -- EV6 84 & Hyundai -- IONIQ 5 84 & \\[3pt]

133 & Audi -- Q6 e-tron Sportback & Audi -- SQ6 e-tron & Audi -- Q6 e-tron quattro \\[3pt]

134 & Porsche -- Macan Turbo Electric & Porsche -- Macan 4 Electric & \\

\end{longtable}
}









\bibliographystyle{elsarticle-num}

\bibliography{Ref}

\end{document}